\documentclass[journal,twoside,web]{ieeecolor}
\usepackage{generic}
\usepackage{amsmath,amssymb,amsfonts}
\usepackage{graphicx}
\usepackage{algorithm}
\usepackage{algpseudocode}
\usepackage{hyperref}
\usepackage{multirow}
\usepackage{soul}

\usepackage[maxnames=1,sorting=none]{biblatex} 
\usepackage{array}
\usepackage{subcaption}

\hypersetup{hidelinks=true}
\usepackage{textcomp}
\def\BibTeX{{\rm B\kern-.05em{\sc i\kern-.025em b}\kern-.08em
    T\kern-.1667em\lower.7ex\hbox{E}\kern-.125emX}}

\title{Cough-E: A multimodal, privacy-preserving cough detection algorithm for the edge}

\begin{document}

\title{Cough-E: A multimodal, privacy-preserving cough detection algorithm for the edge}



\author{Stefano Albini$^{1}$, Lara Orlandic$^{1}$, Jonathan Dan$^{1}$, Jérôme Thevenot$^{1}$, Tomas Teijeiro$^{2}$, Denisa‑Andreea Constantinescu$^{1}$, and David Atienza$^{1}$
\thanks{$^{1}$Embedded Systems Laboratory (ESL) of EPFL, Lausanne, Switzerland. Corresponding authors: \href{stefano.albini@epfl.ch}{stefano.albini@epfl.ch}, \href{lara.orlandic@epfl.ch}{lara.orlandic@epfl.ch}}
\thanks{$^{2}$BCAM - Basque Center for Applied Mathematics, Bilbao, Spain.}
\thanks{This research was supported in part by the Wyss Center for Bio and Neuro Engineering: Lighthouse Noninvasive Neuromodulation of Subcortical Structures, by the ACCESS – AI Chip Center for Emerging Smart Systems, sponsored by InnoHK
funding, Hong Kong SAR, by the EU’s Horizon 2020 grant agreement no. 101017915 (DIGIPREDICT), and in part by the Swiss State Secretariat for Education, Research, and Innovation (SERI) through the SwissChips research project.}
\thanks{T. Teijeiro is supported by the grant RYC2021-032853-I funded by MCIN/AEI/ 10.13039/501100011033 and by European Union NextGenerationEU/PRTR.}
}

\maketitle

\begin{abstract}

Continuous cough monitors can greatly benefit doctors for home monitoring and treatment of respiratory diseases. Although many works propose algorithms to automate this task, they are still limited in terms of poor data privacy and short-term monitoring. Edge-AI is a promising paradigm to overcome these limitations by processing privacy-sensitive data close to their source. However, it presents challenges for the deployment of performant but resource-demanding algorithms on constrained devices.
In this work, we propose a hardware-aware methodology for developing a cough detection algorithm, analyzing design-time trade-offs for performance and energy.
From a suitable selection of audio and kinematic signals, our methodology aims at the optimal selection of features via Recursive Feature Elimination with Cross-Validation (RFECV), which exploits the explainability of the selected XGB model. Additionally, it analyzes the use of Mel spectrogram features, instead of the more common MFCC. Moreover, a set of hyperparameters for a multimodal implementation of the classifier is explored. 
Finally, it evaluates the performance based on clinically relevant event-based metrics.
We apply our methodology to develop Cough-E, an energy-efficient, multimodal and edge AI cough detection algorithm. It exploits audio and kinematic data in two distinct classifiers, jointly cooperating for a balanced energy and performance trade-off.
We demonstrate that our algorithm can be executed in real-time on an ARM Cortex M33 microcontroller.
Cough-E achieves a 70.56\% energy saving when compared to the audio-only approach, at the cost of a 1.26\% relative performance drop, resulting in a 0.78 F1-score.
Both Cough-E and the edge-aware model optimization methodology are publicly available as open-source code.
This approach demonstrates the benefits of the proposed hardware-aware methodology to enable privacy-preserving cough monitors on the edge, paving the way to efficient cough monitoring. 

\end{abstract}



\section{Introduction}
\label{sec:introduction}

Chronic cough is a common symptom associated with a wide range of respiratory conditions. 
It is estimated to affect 5 to 10\% of the global adult population \cite{song2015global}, and has been shown to be the most common cause of doctor visits \cite{finley2018most}. It greatly hinders quality of life by causing chest pain and throat irritation. In severe cases, it can even cause depression and anxiety \cite{chamberlain2015impact, dicpinigaitis2006prevalence, kuzniar2007chronic}.
In recent years, cough has become even more pervasive due to the COVID-19 pandemic. Long COVID-19 has been associated with chronic cough \cite{kanemitsu2024relevant}.

To optimize treatment strategy, physicians would benefit from a quantifiable and accurate measure of patients' cough frequency. 
Precise cough counting could also serve as an indicator for screening of pulmonary diseases which can exhibit cough as one of the first symptoms \cite{kardos2010management}, thus aiding in early treatment \cite{de2011cough}.
Traditionally, cough monitoring has relied on patient self-reporting and intermittent clinical assessments, such as the Leicester Cough Questionnaire (LCQ) and Visual Analog Scale (VAS). However, these methods are moderately correlated with objective cough counts \cite{birring2006cough}. More recently, some clinics have set up monitoring of patients for up to 24 hours \cite{gabaldon2022longitudinal}. Although these setups do capture coughs accurately, they are conducted over a period too short to capture the progression of the cough condition over time.

\begin{figure}[!t]
\centerline{\includegraphics[width=\columnwidth]{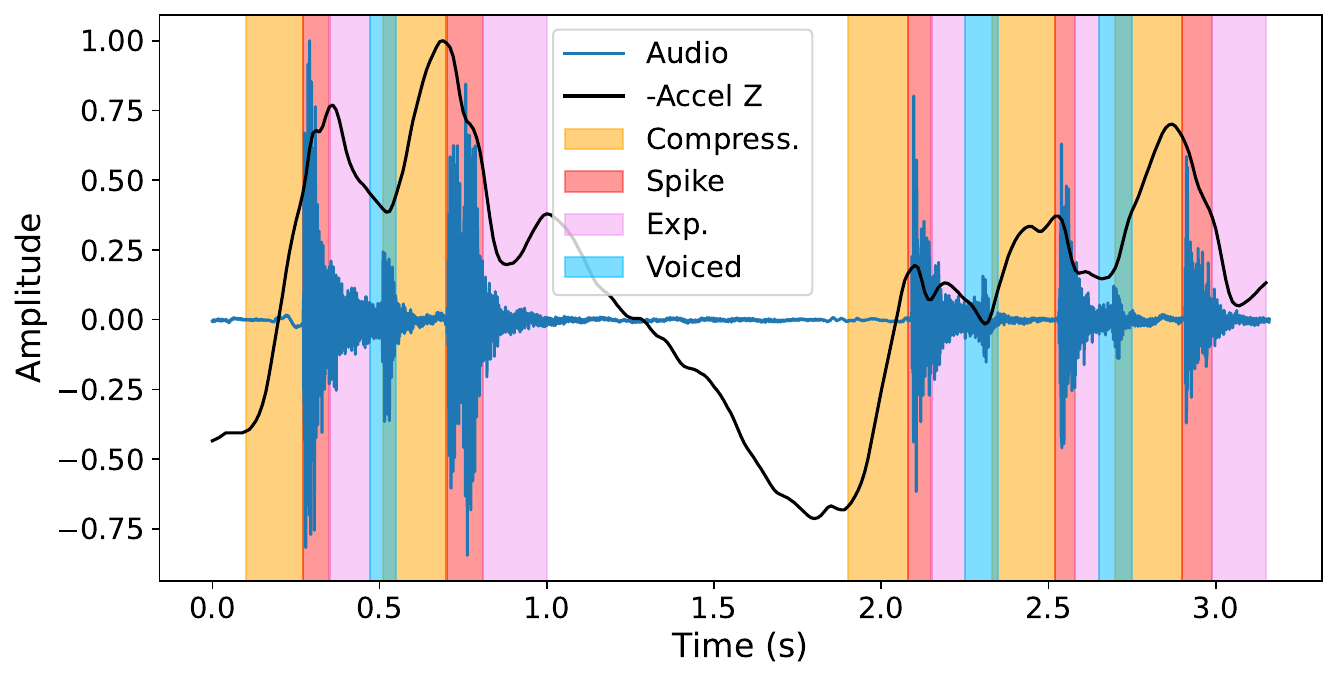}}
\caption{Delineated phases of two bouts of cough events captured by a microphone and an accelerometer placed on the chest of the subject.}
\label{fig:cough_delineation}
\end{figure}
The advent of the Internet of Medical Things (IoMT) poses promising approaches to monitoring patients' healthcare by sensing and processing bio-signals with computerized devices. 
Cough can be measured through audio recordings of vocalization or lung sounds \cite{drugman2020audio, kvapilova2020continuous}, and with accelerometers or gyroscopes capturing chest and neck movements \cite{mohammadi2019automatic, otoshi2021novel}.
Figure \ref{fig:cough_delineation} shows an audio and a kinematic signal (accelerometer measuring sagittal front and back chest movement) during the different phases of a cough, highlighting how differently the two sensors capture the occurrence of a cough.
Some works compared cough-related signals, demonstrating that audio and kinematics are the most performant for automated cough detection. 
Audio sensors have exhibited the highest performance in single-modality systems. Instead, kinematic signals are useful in capturing abrupt chest movements caused by coughs, and can thus aid in distinguishing cough from speech \cite{drugman2013objective}. Furthermore, kinematic signals are unaffected by acoustic background noise and can thus add robustness in noisy environments. However, they can be susceptible to motion artefacts such as walking.

Several commercial products allow reliable continuous recording of audio \cite{urban2022validation,RESP_biosensor, mcguinness2012p159}, and some of these devices also provide automatic cough counting algorithms \cite{birring2008leicester,kuhn2023validation}. 
Similarly, many research works proposed solutions for unobtrusive monitoring with continuous data recording \cite{drugman2020audio, jokic2022tripletcough}, employing smartphone-based applications \cite{kvapilova2020continuous, shim2023smartphone}, fabric-based stretchable-strain sensors \cite{otoshi2021novel, zhang2024high}, a vibroacoustic sensor \cite{mlynczak2015automatic}, arm-secured devices \cite{yang12automatic}, or targeting ear-worn devices \cite{wang2022hearcough, zhang2022coughtrigger, nemati2021coughbuddy}.
Most of these works rely on offline computations, thus requiring data transmission to the cloud or downloading them to a local computer.
This constitutes a privacy issue, particularly for acoustic signals, resulting in bystanders being recorded without their consent. Therefore, the transfer of patients' data poses a significant risk to private data being accessible by a third party.
However, in the context of IoMT, data security of wearable devices can be enhanced using edge Artificial Intelligence (edge AI), a novel computing paradigm where all of the computation occurs on the device itself \cite{singh_edge_2023}.
Edege-AI devices must be small and portable to ensure the user's comfort. Consequently, the embedded microprocessors contain limited computing resources, small memory sizes, and miniature batteries. Furthermore, data must be processed in real-time such that the raw signal can be discarded after classification. This constrains choices of parameters such as sensing modalities, sampling frequencies, and signal window length. If not properly selected, they can hinder energy efficiency, real-time execution, or prevent embedded deployment.

Table \ref{tab:soa_comparison} summarizes some of the related works, showcasing their main figures of comparison. 
In the table, \textit{HW-aware design} refers to the presence of design-time trade-off decisions that consider the impact on measured embedded hardware (HW) metrics, such as energy consumption, memory, and execution time. None of the existing literature is explicitly designed with HW constraints.
Otoshi et al. \cite{otoshi2021novel} are the only ones that did not use audio-based sensors. Rather, they used a 3-axial accelerometer in the epigastric region and a stretchable strain sensor around the neck. 
On the contrary, all the other studies relied on audio signals, either using only microphones \cite{drugman2020audio, wang2022hearcough, jokic2022tripletcough}, or pairing them with kinematic sensors \cite{zhang2022coughtrigger, nemati2021coughbuddy, kuhn2023validation}. 
In the latter case, data from multiple sensors can be jointly fed to a single classifier, or utilized to train separate classifiers that mutually interact.
For instance, some works proposed audio classifiers to trigger a kinematic-based algorithm \cite{nemati2021coughbuddy}, or a kinematic-based classifier that triggers the audio one \cite{zhang2022coughtrigger}, as done in this work.
However, these related works did not explore the cooperation scheme and its effect on runtime, execution, and performance.
Despite this extensive use of audio signals, only some work tackles the issue of data privacy either by transmitting only raw data segments that exceed an amplitude threshold \cite{kuhn2023validation}, or by providing a complete edge AI execution \cite{wang2022hearcough, zhang2022coughtrigger}.
Most works rely on deep networks, such as Convolutional Neural Networks (CNN) or Variational Autoencoders, which are often not suitable for resource-constrained edge devices.
Wang et al. \cite{wang2022hearcough} tackle privacy issues by running on the edge. However, the authors use 4 audio signals, requiring large devices, and no HW-aware design of the algorithms was performed.
Table~\ref{tab:soa_comparison} shows that very few works focused on efficient edge AI execution, analyzing and exploiting algorithmic characteristics and HW-aware trade-offs.

In addition to privacy and implementation concerns, cough detection devices must provide accurate insights into patients' daily cough patterns. To this end, they must be evaluated in terms of clinically relevant performance metrics and tested in everyday noisy environments. Most state of the art (SoA) cough detection algorithms measure their model's performance in terms of sensitivity and specificity at the segment level \cite{otoshi2021novel, drugman2020audio, nemati2021coughbuddy}. However, these metrics do not correspond to the most relevant information for clinicians in evaluating cough severity, such as the precise number of cough events, number and duration of cough bouts, and how many false positive coughs were detected by the wearable device \cite{morice_ers_2007, turner_measuring_2023}. This is in part because algorithms that operate on multiple-second-long windows of data cannot determine how many coughs are present in a segment, nor the temporal distribution of the coughs \cite{otoshi2021novel}. Furthermore, several algorithms are only tested in controlled, noiseless environments that are not representative of real-life scenarios \cite{jokic2022tripletcough, nemati2021coughbuddy}. Lastly, only one work is fully based on open datasets \cite{otoshi2021novel}, and few of them are using a mix of public and custom-collected ones \cite{wang2022hearcough, kuhn2023validation}. Similarly, the code is only made public in \cite{jokic2022tripletcough}, hindering the reproducibility of experiments.

\newcommand*{\crosssymbol}{%
    \text{%
      \raisebox{1ex}{%
        \makebox[0pt][l]{%
          \rule[-.2pt]{.75ex}{.4pt}%
        }%
        \makebox[.75ex]{%
          \rule[-1ex]{.4pt}{1.5ex}%
        }%
      }%
    }%
}   
\newcommand*{\crossupsidedown}{%
    \text{%
      \raisebox{.5ex}{%
        \makebox[0pt][l]{%
          \rule[-.2pt]{.75ex}{.4pt}%
        }%
        \makebox[.75ex]{%
          \rule[-.5ex]{.4pt}{1.5ex}%
        }%
      }%
    }%
}

\begin{table*}[]
\centering
\caption{Comparison of this work with SoA cough monitoring systems}
\begin{tabular}{|c|c|c|c|c|c|c|}
\hline
\textbf{Ref.}          & \textbf{\begin{tabular}[c]{@{}c@{}}Edge-AI\\ execution\end{tabular}} & \textbf{\begin{tabular}[c]{@{}c@{}}HW-aware\\ design\end{tabular}} & \textbf{\begin{tabular}[c]{@{}c@{}}ML\\ performance\end{tabular}}      & \textbf{\begin{tabular}[c]{@{}c@{}}Sensors\end{tabular}}                          & \textbf{\begin{tabular}[c]{@{}c@{}}ML\\ model\end{tabular}}                                     & \textbf{\begin{tabular}[c]{@{}c@{}}Environmental\\ noise\end{tabular}} \\ \hline \hline
Drugman, 2020, \cite{drugman2020audio}      & No  & No  & \begin{tabular}[c]{@{}c@{}}*Sens. = 94.7\% \\ *Spec. = 95.0\% \end{tabular} & \begin{tabular}[c]{@{}c@{}}audio microphone\\ contact microphone\end{tabular}   & \begin{tabular}[c]{@{}c@{}}2 Neural\\ Networks\end{tabular}  & Yes   \\ \hline
Otoshi, 2021, \cite{otoshi2021novel} & No   & No  & \begin{tabular}[c]{@{}c@{}}*Sens. = 92.0\% \\ *Spec. = 96.0\% \end{tabular} & \begin{tabular}[c]{@{}c@{}}3-axis accelerometer\\ stretchable strain sensor\end{tabular} & \begin{tabular}[c]{@{}c@{}}Variational Autoencoder\\ +\\ K-means clustering\end{tabular}   & Yes  \\ \hline
Nemati, 2021, \cite{nemati2021coughbuddy}  & No & No & \begin{tabular}[c]{@{}c@{}}*Sens. = 83.0\% \\ *Spec. = 91.7\% \end{tabular} & \begin{tabular}[c]{@{}c@{}}microphone\\ kinematic\end{tabular}  & \begin{tabular}[c]{@{}c@{}}Random forest\\ +\\ DWT template matching\end{tabular} & No \\ \hline
Wang, 2022, \cite{wang2022hearcough}  & Yes   & No  & \begin{tabular}[c]{@{}c@{}}*Acc. = 78.5\% \\ *F1 = 77.0\%*\end{tabular}  & 4 microphones   & CNN  & Yes \\ \hline
Jokic, 2022, \cite{jokic2022tripletcough}  & No & No  & *Acc. = 80.0\% & 4 microphones  & CNN & No\\ \hline
Zhang, 2022, \cite{zhang2022coughtrigger}  & Yes & No & *AUROC = 77.0\% & \begin{tabular}[c]{@{}c@{}}microphone\\ kinematic\end{tabular} & \begin{tabular}[c]{@{}c@{}}Multi-Nearest Center\\ Classifier\end{tabular} & Yes  \\ \hline
Kuhn, 2023, \cite{kuhn2023validation}  & No & No & \begin{tabular}[c]{@{}c@{}}*Sens. = 88.5\% $\dagger$  / 84.15\% \crossupsidedown \\\ *Prec. = 99.97\% $\dagger$  / 99.7\% \crossupsidedown \end{tabular} & \begin{tabular}[c]{@{}c@{}}microphone\\ vibrations sensor\end{tabular} & DNN & Yes  \\ \hline
\textbf{Cough-E (ours)}   & Yes  & Yes  & \begin{tabular}[c]{@{}c@{}}Sens. = 71.0\% \\ Prec. = 78.0\%\\ F1 = 77.7\%\end{tabular} & \begin{tabular}[c]{@{}c@{}}microphone\\ kinematic\end{tabular} & \begin{tabular}[c]{@{}c@{}}2 XGB\\ classifiers\end{tabular}  & Yes  \\ \hline 
\multicolumn{4}{c}{(*)Segment-based performance metrics (cf. Section \ref{exp_setup:sub:ml_performance})} & 
\multicolumn{2}{c}{\begin{tabular}[c]{@{}c@{}} ($\dagger$) daytime testing \\ (\crossupsidedown) nightime testing\end{tabular}}
\end{tabular}
\label{tab:soa_comparison}
\end{table*}




In this work, we focus on the design of an automated cough monitor. We propose a methodology to jointly tackle cough classification and edge-AI challenges.
Particularly, we compare cough detection classifiers based on audio and kinematic signals, as well as a combination of both, evaluating them on event-based metrics. Several hyperparameters, such as sampling frequency, window length, and number of features, are compared and analyzed on embedded metrics (i.e. energy, memory, duty cycle),
This design trade-off analysis maximizes cough detection performance metrics while minimizing embedded ones, enabling real-time edge-AI execution.  
With these premises, the main contributions of our works are the following:
\begin{itemize}
    \item We develop a methodology that embeds hardware-aware decisions into the main phases of the machine learning development pipeline. This allows the embedded execution of an audio-based model on a low-power microcontroller for a 3.65\% reduction in cough detection performance.
    \item We apply the methodology to develop Cough-E, a real-time, edge-AI cough monitoring application leveraging the strengths of two sensing modalities. This novel algorithm reduces energy consumption by 70.56\% for only a 1. 26\% reduction in the F-1 score compared to the optimized audio-based model.
    \item We demonstrate the efficiency of using the raw Mel spectrogram feature instead of MFCCs, exhibiting higher separability and providing better performances while achieving a 20x energy reduction.
    \item We provide open code for the methodology, as well as the embedded version of our optimized application, available here: \url{https://github.com/esl-epfl/Cough-E}
\end{itemize}


\section{Methodology}
\label{sec:methodology}

\begin{figure*}[!t]
\centerline{\includegraphics[width=\textwidth]{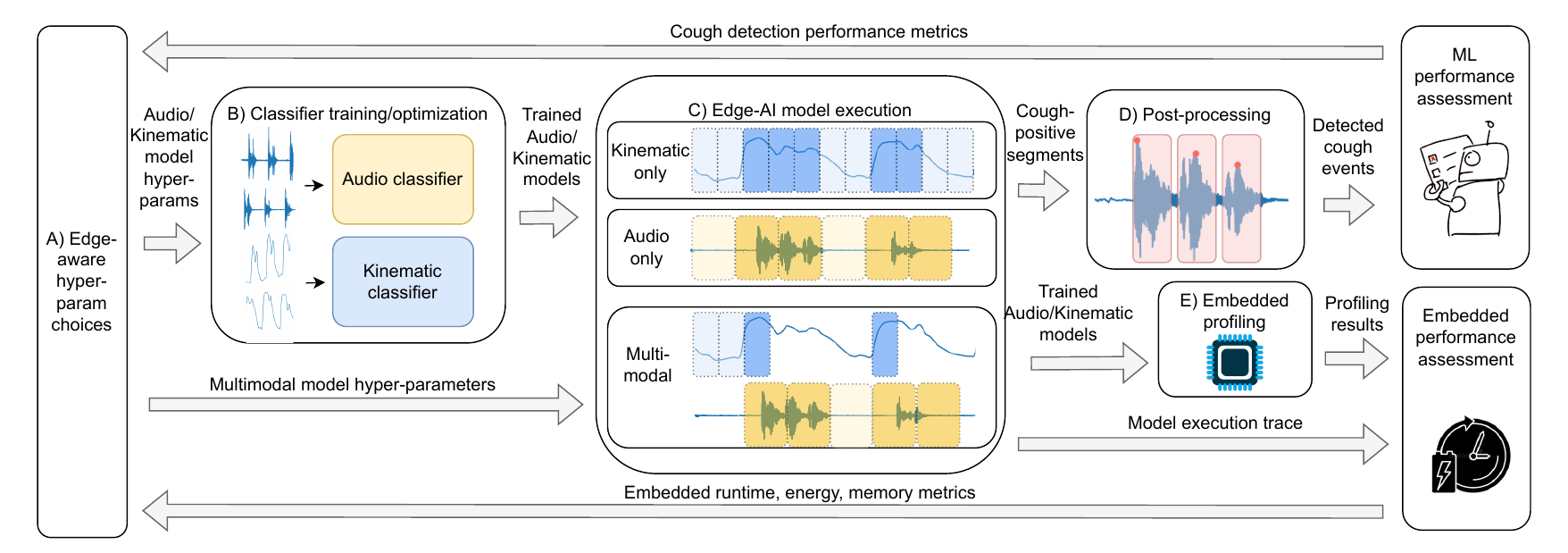}}
\caption{Overview of the methodology proposing an energy-efficient, real-time, edge-AI cough detection model}
\label{methodology:fig:methodology_overview}
\end{figure*}

This paper introduces a methodology for the development of an energy-efficient, accurate and edge-aware cough detection model. The methodology and organization of this section are illustrated in Figure~\ref{methodology:fig:methodology_overview}. First, in Section \ref{methodology:sub:hyperparameters}, we detail the hyperparameters that affect both machine learning (ML) and embedded performance metrics. In Section \ref{methodology:sub:ml_training}, we present the training procedure used to develop lightweight, explainable classifiers. Then, in Section \ref{methodology:sub:simulator}, we explain how each classifier is run across testing signals to mimic edge-AI execution. Section \ref{methodology:sub:multimodal} details the design of our novel multi-modal algorithm, Cough-E, that leverages both the audio and kinematic classifiers. Section \ref{methodology:sub:postprocessing} describes our physiology-inspired post-processing method to exact start and endpoints of cough events. Then, in Section \ref{exp_setup:sub:ml_performance}, we describe how these predicted cough regions are compared to the ground truth to obtain fine-grained, clinically useful performance metrics.

\subsection{Edge-AI hyperparameter trade-off analysis}
\label{methodology:sub:hyperparameters}
Many design choices made prior to model training affect the model's ability to distinguish coughs from non-coughs, as well as the embedded runtime, energy consumption, and memory footprint. In this section, we outline the most significant hyperparameters that need to be co-optimized for both ML and embedded performance.

\subsubsection{Signal selection}
\label{methodology:sub:hyperparameters:signals}
 We analyze the public data from Orlandic et al.  (described in Section \ref{exp_setup/dataset}, i.e. audio and kinematic modalities~\cite{orlandic2023multimodal}). 
We evaluate whether classification performance is more robust using audio signals from chest- or outward-facing microphones placed on the chest.
We utilize kinematics from an inertial measurement unit (IMU) also placed on the chest, specifically tri-axial accelerations and derived 3D angles (Yaw-Y, Pitch-P, Roll-R) from the manufacturer data fusion algorithm.

Classification performance is evaluated in terms of average F-1 score over cross-validation (CV) folds.

\subsubsection{Signal sampling rate}
\label{methodology:sub:hyperparameters:sampling}

We investigate the effects of sampling rate on audio-based classification. We test the difference in performance between signals re-sampled at 16, 8, and 4 kHz by applying low-pass filtering and decimation.

\subsubsection{Window length}
\label{methodology:sub:hyperparameter:window_length}
We evaluate the performance of the classifier as a function of the input window length.
Since cough events typically last up to 0.5 seconds \cite{chang2006physiology}, we evaluate window lengths in the range [0.4, 1.0] seconds in increments of 0.1 seconds.

\subsubsection{Feature extraction and optimization}
\label{methodology:sub:hyperparameters:feature_ext}



\begin{table}[]
\caption{Features extracted from each signal}
\centering
\begin{tabular}{|c|c|c|c|}
\hline
\textbf{Signals}                                                                                        & \begin{tabular}[c]{@{}c@{}}\textbf{Feature}\\ \textbf{count}\end{tabular}     & \begin{tabular}[c]{@{}c@{}}\textbf{Feature}\\ \textbf{type}\end{tabular}       & \begin{tabular}[c]{@{}c@{}}\textbf{Feature} \\ \textbf{list}\end{tabular}                                                                                                                                   \\ \hline
\multirow{3}{*}[-2.5em]{\begin{tabular}[c]{@{}c@{}} Microphone \end{tabular}}   & 14                                                                       & \begin{tabular}[c]{@{}c@{}}Freq.\\ domain\end{tabular}       & \begin{tabular}[c]{@{}c@{}}Spectral decrease/ \\ slope/ roll-off/ skew/\\ centroid/ spread/ \\ flatness/ st. dev/\\ entropy, Dominant\\ frequency, Power \\ spectral density\end{tabular} \\ \cline{2-4} 
                                                                                               & \begin{tabular}[c]{@{}c@{}}52 (MFCC)\\ or 256 \\ (Mel Spec)\end{tabular} & \begin{tabular}[c]{@{}c@{}}Mel\\ freq.\\ domain\end{tabular} & \begin{tabular}[c]{@{}c@{}}MFCC or Mel Spec.\\ mean/ st. dev/ max/ \\ entropy\end{tabular}                                                                                                \\ \cline{2-4} 
                                                                                               & 22                                                                       & \begin{tabular}[c]{@{}c@{}}Time\\ domain\end{tabular}        & \begin{tabular}[c]{@{}c@{}}Root mean square,\\ Zero-crossing rate,\\ Crest factor,\\ Energy envelope peak\\ detection (EEPD)\end{tabular}                                                       \\ \hline
\begin{tabular}[c]{@{}c@{}}Accel X, Y, Z,\\ Accel norm,\\ Angles Y, P, R,\\ Angles norm\end{tabular} & 13                                                                       & \begin{tabular}[c]{@{}c@{}}Time \\ domain\end{tabular}       & \begin{tabular}[c]{@{}c@{}}Zero-crossing rate,\\ Root mean square,\\ Crest factor, Kurtosis,\\ Line length, Approx.\\ zero crossing\end{tabular}                                         \\ \hline
Subject characteristic                                                                                        & 2                                                                        & Constant                                                      & Gender, BMI                                                                                                                                                                               \\ \hline
\end{tabular}
\label{tab:extracted_features}
\end{table}
The features we use are listed in Figure \ref{tab:extracted_features}. Along with the audio and kinematic features,  gender and BMI are also provided. The audio features are taken from prior work \cite{orlandic_coughvid_2021, orlandic2023multimodal}. 

\begin{figure}[!t]
\centerline{\includegraphics[width=\columnwidth]{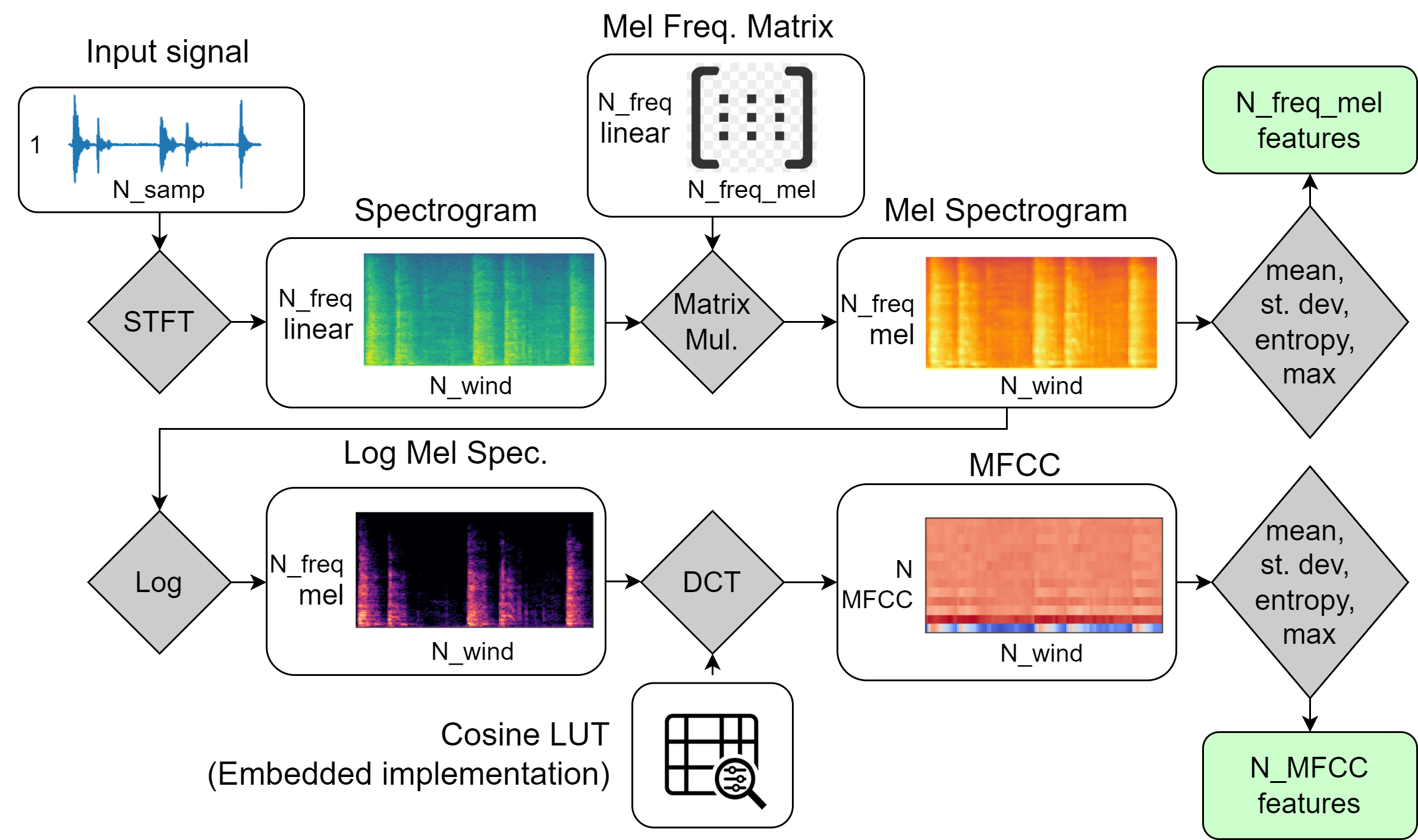}}
\caption{Computation of Mel Spectrogram and MFCCs listing the inputs and dimensions at each step.}
\label{fig:mfcc_decomposition}
\end{figure}

The MFCC computation is depicted in Fig. \ref{fig:mfcc_decomposition}; the input signal is first split into windows and a spectrogram is computed (Short-time Fourier Transform, STFT). Next, a matrix multiplication transforms the spectrogram 
into the logarithmic Mel domain, which has 64 frequency bands (i.e. $N_{freq\_mel}=64$). 
Furthermore, a logarithm is taken and Discrete Cosine Transform (DCT) is performed using a Look-Up table (LUT) of precomputed cosine values, preloaded into the embedded implementation of the application.
This produces the final 13 MFCCs ($N_{MFCC} = 13$). In this work, we investigate whether it is necessary to perform the arithmetically complex logarithm and DCT steps to obtain MFCCs, or whether spectral information from the Mel spectrogram is enough to distinguish between cough and non-cough samples. As a result, four statistical metrics -- namely, mean, standard deviation, entropy, and maximum -- are computed per window to obtain 256 Mel spectrogram or 52 MFCC features, respectively. 

For the kinematic classifier, each signal is analyzed separately, as well as the $\ell_2$ vector norms of each type of signal (i.e. accelerations and angles). The same 13 time-domain features are computed for each signal. We use common features found in cough counting classifiers from kinematic signals \cite{orlandic2023multimodal} and we add a SoA feature previously used for real-time EEG signal processing: Approximate Zero Crossings (AZC) \cite{zanetti_approximate_2022}. The AZC feature estimates the dominant frequency in a low-complexity fashion. Therefore, we extracted this feature from the kinematic signals for the first time to detect abrupt slopes at the start of cough events.
To compute AZC, the segment is first simplified using the Douglas–Peucker method as a polygonal approximation algorithm \cite{douglas1973algorithms}. This method outputs a subset of the initial samples, interpolating them based on an amplitude threshold $\epsilon$. This operation extracts the most abrupt changes in the signal while removing noise (the higher the tolerance $\epsilon$, the simpler the output signal). Finally, a zero-crossing count is implemented on the simplified signal as an estimation of the dominant frequency.
In our implementation, we explore a set of $\epsilon$ values between 0.3 and 1.0 in increments of 0.1 to capture a higher set of approximate frequencies.

\subsubsection{Feature selection}
\label{methodology:sub:hyperparameters:feature_sel}
For each feature in Table \ref{tab:extracted_features}, the computational demand vs. the performance contribution is assessed. We use Recursive Feature Elimination with Cross-Validation (RFECV) to estimate the performance of the ML classifier with a reduced feature set \cite{awad_recursive_2023}. The full feature set is validated in a nested CV loop, and after each test, the 10 features with the lowest importance scores are eliminated and the process is repeated. As we use tree-based classifiers, the importance score is defined as the number of times a given feature is used to split the data between all trees \cite{noauthor_xgboost}. Concurrently, we analyze the duty cycle of each feature set on the embedded platform described in Section \ref{exp_setup/embed_c}. Using both metrics, we determine the optimal number of features to balance the trade-off between computational efficiency and classification performance.

\subsection{Classifier training and optimization}
\label{methodology:sub:ml_training} 
Once a set of hyperparameters is defined, the next step in our edge-AI model development pipeline is to train lightweight, explainable ML classifiers using audio and kinematic modalities. For the kinematic classifier, all available signals are used to train a single model, whereas different models are trained for the inner and outer audio signals. The model training, validation, and testing methodology is explained in Section \ref{exp_setup/ml_test_train}.

First, the signals in the training dataset are segmented with no overlap. A segment is labeled as cough-positive if either the majority of the segment's duration contains a ground-truth cough, or the majority of at least one ground-truth cough is contained within the segment. This accounts for both scenarios in which the window length is close to the length of a cough, or long windows that contain multiple cough sounds.

Next, features (cf. Table \ref{tab:extracted_features}) are extracted for the audio and kinematic classifiers. The kinematic classifier additionally uses gender and BMI metadata, as both can impact the subject's body shape and consequent motions, whereas the audio classifier only considers the gender feature. 

To enhance the generalizability of the audio model to unseen test subjects, the training set is augmented in a semi-supervised fashion using data from the COUGHVID dataset \cite{orlandic_coughvid_2021}, as performed in \cite{orlandic_semi-supervised_2023}. To this end, a preliminary classifier is trained using the data described in Section \ref{exp_setup/dataset}. This classifier is then tested on randomly selected recordings of the COUGHVID dataset, which contains cough recordings, but no fine-grained labeling of which signal segments contain coughs. If the classifier detects a cough segment, the training dataset is augmented with the features of that segment. This process is repeated until the training dataset has an equal number of cough and non-cough segments.

The feature vectors and their corresponding labels are subsequently used to train an eXtreme Gradient Boosting (XGB) model. Next, RFECV is performed to analyze the effects of reducing the feature set on the processing overhead and ML performance of the classifier. RFECV is performed in a nested CV fashion; within each Leave-One-Subject-Out (LOSO) fold, an inner 5-fold CV is performed to assess the performance of each reduced feature set. Evaluating this trade-off across the outer CV folds results in a choice of the optimal number of features, $n$.  Then, the feature set is reduced by averaging the feature importance 
(i.e. the  percentage of times a feature is used to split the data across all trees)
across the inner CV folds and retaining the $n$ features with the highest average importance scores. This step enhances the embedded efficiency of the model execution by reducing unnecessary feature computation. Finally, a classifier is trained for each CV fold using the reduced feature set.

\subsection{Edge-AI Model Execution}
\label{methodology:sub:simulator}

Edge AI execution is performed as a sliding window operation.
After one window is sampled and buffered, the corresponding model starts its execution.
For both models, subsequent windows have a certain overlap, which affects the real-time deadline.
Therefore, the computations are not continuous with the stream of incoming data, but require each time a full window in memory. After the execution, this data is not used anymore and can be discarded, hence saving memory and ensuring privacy.

\subsection{Multimodal model design and execution}
\label{methodology:sub:multimodal}
In addition to our methodology for optimizing audio-based and kinematic-based ML models, we propose Cough-E, a novel co-operative algorithm that leverages the strengths of both modalities. The audio-based cough detection models tend to exhibit higher accuracy than their kinematic-based counterparts \cite{drugman2013objective}, but at the cost of up to 160x the sampling frequency and consequent processing overhead. This work proposes a multimodal classification approach that utilizes kinematic classifier as a triggering mechanism for the audio classifier.
With this scheme, the kinematic model runs until it detects a cough, and then activates the audio model to verify the classification. The goal of this multimodal scheme is to minimize loss in performance and incur energy savings.

\begin{figure}[ht]
    \centering
    \begin{subfigure}[b]{\columnwidth}
        \centering
        \includegraphics[width=\columnwidth]{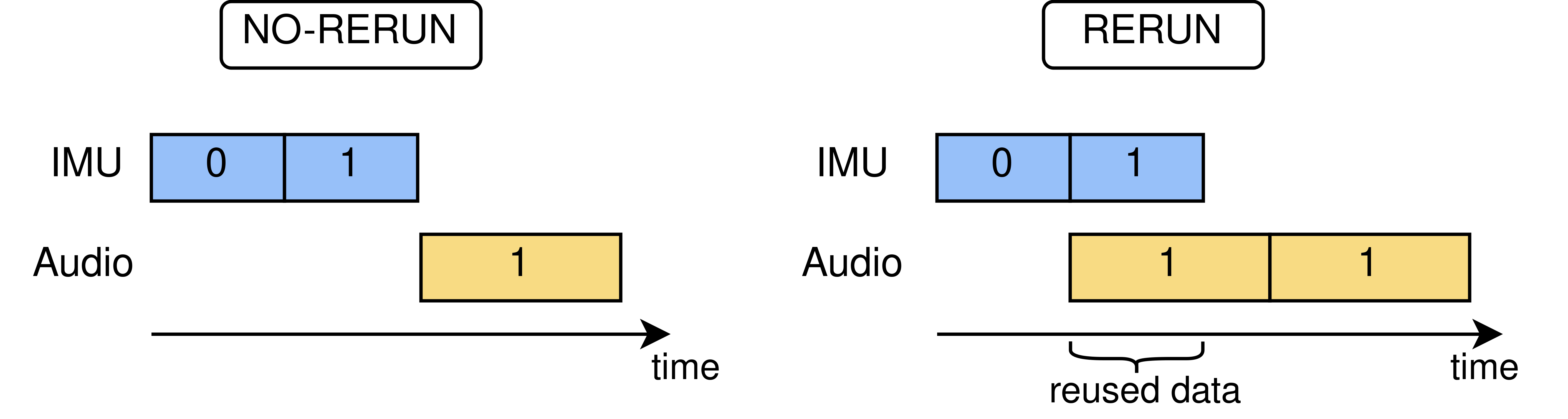}
        \caption{Example of the alternation scheme of the two models in \textit{no-rerun} (left) and \textit{rerun} (right) modes of operation.}
        \label{fig:rerun_no_rerun_flows}
    \end{subfigure}
    
    \vspace{0.5cm}
    
    \begin{subfigure}[b]{\columnwidth}
        \centering
        \includegraphics[width=\columnwidth]{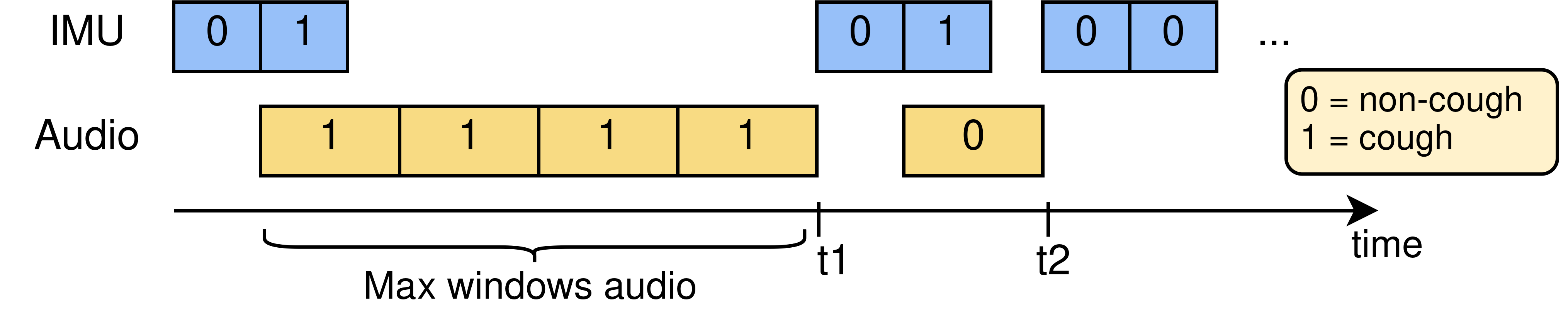}
        \caption{Example of full cooperation in \textit{rerun} mode.}
        \label{fig:cooperative_scheme}
    \end{subfigure}
    
    \caption{Schematic cooperation flows for the two models (top) and example of cooperation (bottom).}
    \label{fig:combined_images}
\end{figure}

The multimodal execution begins with the kinematic model running continuously. Once the kinematic classifier detects a cough, it halts, and the audio classifier begins its execution. 
When triggered, the audio-model can start its execution according to one of two modes of operation:
\begin{itemize}
    \item \textit{no-rerun}: the audio model starts its execution from the next window after the one that caused a positive classification in the kinematic model (Figure \ref{fig:rerun_no_rerun_flows}, left).
    \item \textit{rerun}: the audio model processes data from the time-window that caused a \textit{cough} classification in the kinematic model, thus reassessing it. In this way, possible errors due to the high sensitivity of the kinematic model can be corrected (Figure \ref{fig:rerun_no_rerun_flows}, right).
\end{itemize}

When triggered, the audio model must run long enough to capture the entire cough bout, but should switch back as soon as possible to minimize the system's energy usage. Therefore, the audio classifier continues until either 
1) it runs for the maximum number of consecutive windows, $N^{windows}_{max}$ (Figure~\ref{fig:cooperative_scheme}, instant $t_1$), or
2) it outputs a non-cough label (Figure~\ref{fig:cooperative_scheme}, instant $t_2$).
The first case prevents edge cases in which the system stays in an energy-demanding execution unnecessarily long due to false-positives, such as noisy environments or bystander coughs. Once either of these conditions is met, the kinematic model resumes its execution.

To enhance both the ML performance and embedded efficiency of Cough-E, several additional hyperparameters are co-optimized:
\begin{itemize}
    \item The choice of \textit{rerun} versus \textit{no-rerun} modes.
    \item The value of the $N^{windows}_{max}$ variable.
    \item The decision thresholds above which the kinematic and audio models classify a sample as a cough, thus triggering model switching.
\end{itemize}

\subsection{Post-processing algorithm}
\label{methodology:sub:postprocessing}
The edge-AI model execution is designed to identify regions of the signal that contain at least a fraction of a cough. However,  the number of windows containing coughs does not accurately quantify individual cough events and does not provide the exact timing of coughs. This information is crucial for clinicians to analyze the temporal pattern of cough events, (i.e. whether single coughs occur sporadically over time or in sudden bouts of multiple coughs), which indicates disease severity and underlying physiological mechanisms \cite{turner_measuring_2023, dockry_s17_2022}. Therefore, to extract the exact start and stop locations of each cough event, we designed a novel audio post-processing algorithm for the audio and multimodal classifiers that is based on cough physiology.

The post-processing algorithm is triggered when the audio classifier detects a cough in a segment. The signal segment is first down-sampled to 2 kHz to reduce processing overhead. It then extracts the peaks and rough start and end locations of the cough segments using hysteresis thresholding on the signal power computed for the segment duration. The start of a cough segment is the point at which the signal goes above the upper threshold, set halfway between the root-mean-square (RMS) and maximum powers of the signal. Then, the end is the location when the signal drops below the lower threshold, set to the RMS power. A cough peak is extracted as the maximum power between the start and end locations. At this point, the raw audio of the signal can be discarded and only the starts, ends, and peaks are retained, thus being consistent with edge-AI execution.

The segment thresholding detects regions of loud bursts, but these do not always correspond to the exact cough onsets and offsets, as they mainly pick up only the explosive cough spike phase. Furthermore, window overlaps mean that some cough regions may be counted twice. Therefore, our cough region refinement algorithm aggregates the peaks of consecutive cough-positive windows and leverages cough physiology and subject-specific information to determine the start and stop locations of each cough event.

Fig.~\ref{fig:cough_delineation} depicts the various phases of a cough and their manifestations in the audio and accelerometer z-direction  -- normal to the chest -- signals. It depicts two cough bouts, defined as successive coughs that occur without pause \cite{dockry_s17_2022}. The timings of the various phases are established in the medical literature \cite{chang2006physiology}. First, the compressive phase (approximately 0.2 s) is characterized by pressure accumulation in the lungs and consequent chest acceleration inward. Next, opening the glottis causes the cough spike phase (0.03-0.05 s) and produces the loudest cough noise. After this, air rushes out of the lungs in the expiratory phase (0.2-0.5 s). Finally, some coughs are followed by a voiced component of unknown duration, which causes another peak in the audio signal that may overlap with the compressive phase of the next cough.

\begin{algorithm}[b]
	\caption{Cough region refinement algorithm}
	\begin{algorithmic}[1]
		\Function{RefineRegions}{${\mathbf{t^{starts}}},{\mathbf{t^{ends}}},{\mathbf{t^{pks}}},{\mathbf{A}}$} 
        \vspace{0.25em}
    	\For{$n = 0$ : $length(\mathbf{t_{peaks}})-2$} \label{alg:lin2}
            \vspace{0.25em}
        	\If{$t^{pks}_{n+1} - t^{pks}_{n} < t_{min}^{coughDur}$} \label{alg:lin3}
                \vspace{0.25em}
        		\State $discardShorterPeak(t^{pks}_{n+1},t^{pks}_{n},A_{n+1}, A_n);$ \label{alg:lin4}
                \vspace{0.25em}
                \State $mergeRegions(t^{starts}_{n+1},t^{starts}_{n},t^{ends}_{n+1},t^{ends}_{n});$ \label{alg:lin5}
            \EndIf \label{alg:lin6}
        \EndFor \label{alg:lin7}
        \vspace{0.25em}
        \State $t^{pkToEnd}_{avg} = avgPkToEndDist(\mathbf{t^{pks}});$ \label{alg:lin8}
        \vspace{0.25em}
        \For{$n = 0$ : $length(\mathbf{t_{peaks}})-1$} \label{alg:lin9}
            \vspace{0.25em}
            \If{$t^{pks}_{n}-t^{starts}_n < t_{min}^{beforePk}$} \label{alg:lin10}
                \vspace{0.25em}
                \State $t^{starts}_n = t^{pks}_n - t_{min}^{beforePk};$\label{alg:lin11} \Comment{Refine start}
            \EndIf \label{alg:lin12}
            \vspace{0.25em}
            \If{$t^{pks}_{n+1} - t^{pks}_{n} < t_{max}^{coughDur}$} \label{alg:lin13}
                \vspace{0.25em}
                \State $t^{ends}_n = t^{starts}_{n+1};$ \Comment{Cough in bout} \label{alg:lin14}
            \Else  \label{alg:lin15}
                \State $t^{ends}_n = t^{pks}_n + t^{cough}_{avg}*C;$ \label{alg:lin16}\Comment{Single cough}
            \EndIf \label{alg:lin17}
        \EndFor \label{alg:lin18}
        \EndFunction 
	\end{algorithmic}
 \label{algo:cough_refinement}
\end{algorithm}

Using this information, we develop Algorithm \ref{algo:cough_refinement} to delineate the starts and stops of each cough event to compare to the ground-truth labels. First, in lines \ref{alg:lin2}-\ref{alg:lin7}, the distance between subsequent peaks is computed to determine whether two peaks belong to the same cough. This can happen if the voiced component of the cough makes a loud noise, or if the signal overlap causes a peak to be counted twice. Peaks are considered part of the same cough if they are closer than $t_{min}^{coughDur}$ apart, which is set to the sum of the minimum cough spike and expiration durations (0.23 s), as these two phases and the optional voiced phase produce the noise that can be detected in the audio signal. If two detected peaks belong to the same cough, the peak with the higher amplitude $A$ is retained and the union of the two cough regions is taken.

Then, in Algorithm \ref{algo:cough_refinement} line \ref{alg:lin8}, $t^{pkToEnd}_{avg}$, the subject-specific average duration from the cough peak to the end of the cough sound, is calculated to help determine the end locations of single coughs or final coughs in a bout. These are otherwise difficult to measure with audio thresholding because they are characterized by a gradual decrease in audio signal amplitude. To compute this metric, the distance between subsequent peaks is computed to determine which coughs are part of a bout by checking if peaks are closer together than $t_{max}^{coughDur}$, set to the maximum possible spike and expiration phase sum (0.55s). Then, the average distance between peaks in a bout is computed, and the minimum possible cough spike duration is subtracted to get the average peak-to-end time. 

Finally, in Algorithm lines \ref{alg:lin9}-\ref{alg:lin18}, the start and stop locations are refined to ensure that they are physiologically plausible and take into account all of the cough phases, not just the cough spike. First, if the start locations are closer than $t_{min}^{beforePk}$, half of the minimum cough spike duration, to the peaks, they are set to the peak minus $t_{min}^{beforePk}$. Next, if the cough is followed by other coughs in a bout (i.e. less than $t_{max}^{coughDur}$ from the following cough), its end location is set to the start location of the following cough. Otherwise, in the case of single coughs or coughs at the end of a bout, the end location is set to its peak plus $t^{pkToEnd}_{avg}$ times the variable $C$, which is an exponentially decreasing constant depending on how many coughs have previously occurred in the bout, as it is observed that coughs decrease in duration as more coughs occur and the lungs empty.

\subsection{ML performance metrics}
\label{exp_setup:sub:ml_performance} 
As proposed in \cite{orlandic_how_2024}, we measure the performance of our classifier in terms of event-based success metrics. Moreover, a True Positive (TP) is counted when the start and end of a predicted cough event overlap with a ground-truth event within 0.25 s tolerance. This value corresponds to the sums of the cough spike and the minimum durations of the compressive / expiration phase, described in Section \ref{methodology:sub:postprocessing}, to account for chest compression prior to a detected audio event or vocal artifacts following a cough. Otherwise, if the detected cough event does not overlap or overlap outside of these tolerance windows, a False Positive (FP) is counted. False Negatives (FNs) are ground-truth events that do not overlap within this tolerance window with any predicted event. Using these event-based measures, we extract the sensitivity (SE), precision (PR), F-1 score (F-1), and false positives per hour (FP/Hr) of each tested classifier. This allows us to quantify model performance in terms of metrics that would be relevant in a real patient monitoring scenario.


\section{Experimental setup}

This section presents the design of the experiments aimed at validating the models' performance and the energy and accuracy trade-off yielded by their joint use on the edge.
In particular, subsection \ref{exp_setup/dataset} details the chosen dataset and its main characteristics pertinent to the algorithm design and validation.
Subsection \ref{exp_setup/ml_test_train} explains how the ML training, validation, and testing were performed to ensure a fair comparison between models and generalizability to unseen test subjects.
In subsection \ref{exp_setup/simulator}, the simulator's experiments are explained, detailing the studied configurations and the hyperparameter tuning.
Finally, subsection \ref{exp_setup/embed_c} provides details on the embedded \texttt{C} implementation of the models, their collaborative scheme, and the chosen wearable device.

\subsection{Dataset}
\label{exp_setup/dataset}
For developing and testing our edge-AI cough detection models, we use the multimodal dataset presented in \cite{orlandic2023multimodal}. This is one of the very few public datasets that provide the fine-grained cough location ground-truth labels necessary for validating cough quantification algorithms.
The data was acquired by a chest-worn wearable device containing two microphones (SPH0645LM4H-B, Knowles, Itasca, USA), one facing toward the chest and one facing away from it, and a 9-axis kinematic (BNO080, Hillcrest Labs, Rockville, USA). The audio and kinematic sensors were sampled at 16 kHz and 100 Hz, respectively.

Experiments were conducted on 20 healthy subjects. The data of 15 subjects, comprising 4 hours of biosignals and more than 4,000 annotated cough events, is used to select hyper-parameters and train the models. The remaining 5 subjects are set aside as a private test set. Subjects were instructed to produce different sounds in varying degrees of acoustic and kinematic noise. The database contains recordings of five distinct sound categories: coughing, laughing, throat clearing, breathing, and talking. The database contains equal numbers of the five sound category recordings to test the ability of the model to distinguish cough from other subject-produced sounds.

The dataset also contains recordings captured in two kinematic noise scenarios (sitting and walking), as well as audio noise scenarios (silence, traffic, music, and bystander coughing). These noise scenarios are useful in evaluating the robustness of models to real-life noise situations that a subject may be exposed to in everyday monitoring.

\subsection{Machine Learning Training and Testing}
\label{exp_setup/ml_test_train}

The entire ML development pipeline follows a nested cross-validation (CV) framework to validate the choices of hyperparameters listed in Sections \ref{methodology:sub:hyperparameters} and \ref{methodology:sub:simulator}. Within this framework, the training dataset described in Section \ref{exp_setup/dataset} is split using LOSO, where the 14 training subjects are used to train the ML classifiers (cf. Section \ref{methodology:sub:ml_training}), then these classifiers are run across the CV test subject emulating edge-AI execution (cf. Section \ref{methodology:sub:simulator}). Within these 14 training subjects, several model optimizations (i.e. RFECV) are made using a nested 5-fold CV and averaging the performance results of each fold. The Average Precision metric \cite{noauthor_average_precision_score_nodate}, which weighs the sensitivity and precision of the model across different decision thresholds, is used as an intermediate, segment-based performance metric for the inner CV folds. For the outer CV folds, the more meaningful event-based ML performance metrics detailed in Section \ref{exp_setup:sub:ml_performance} are computed and averaged across the 15 folds. Finally, to ensure the generalizability of our final edge AI models, five unseen test subjects are used to evaluate the success of the final trained edge-AI models. For the audio, kinematic, and multimodal models, the final model is trained using the full 15 training subjects and the hyperparameters selected through nested CV. This model is then run in an edge-AI fashion on each of the testing subjects only once to provide the final testing results.

\subsection{Simulator}
\label{exp_setup/simulator}
To evaluate the impact of multimodal parameters on embedded metrics such as execution time and energy consumption, we develop a Python simulator that replicates the real-time edge-AI execution flow of the two cooperating models.
The simulator supports three execution configurations: kinematic model only, audio model only, and multimodal. 
This allows us to explore the impact of the models' cooperation and the multimodal hyperparameters, discussed in Section \ref{methodology:sub:multimodal}, on energy, runtime, and performance when compared to the single-model executions.

Each simulation generated an output file containing the hyperparameters setting and an execution trace.
The trace is composed of a sequence of labels that indicate which models have been executed in a specific window, providing information on the frequency of executions of one model versus the other.
The output file is processed by a custom-made Python script to estimate the full energy consumption and execution time of the given simulation.
To do this, it utilizes the information of the trace on when each model is executed, combined with off-line energy and runtime measurements of the corresponding model.

\subsection{Embedded code development and profiling}
\label{exp_setup/embed_c}

To properly provide embedded parameters of the models to the simulator (cf. Section \ref{exp_setup/simulator}),  energy and time measurements are performed per each model. 
With this goal, Cough-E is deployed on a SoA microcontroller, selected as a use case to provide real energy measurements applied to a realistic deployment scenario.

The embedded application is developed in the \texttt{C} language with 32-bit floating point representation.
The majority of the feature extraction and ML routines are custom-made, except for the FFT algorithm for which the \textit{kiss\_fft} library \cite{kiss_fft_library} is used with pre-computed twiddle factors to avoid donline computational overhead.
Similar offline pre-calculations are used for other static data, such as the Mel basis for the Mel spectrogram computation, and the Hanning windows for the Short-Time-Fourier-Transform (STFT) and Power-Spectral-Density (PSD) algorithms.
Similarly, the weights of the two classifiers are saved in custom-made LUTs.
These pre-computed data constitute static memory constraints of the application, imposing memory requirements on the embedded hardware to select. Moreover, this pre-computation enables saving computation time and energy at the cost of higher memory usage.
Lastly, cough peaks, starts, and ends are extracted per each window (cf. Section \ref{methodology:sub:postprocessing}). Algorithm \ref{algo:cough_refinement} is then applied every 5 seconds to refine the obtained cough regions and provide a final cough count.

In the translation process from Python to \texttt{C}, various optimizations are performed. Particularly, the computation of the Mel spectrogram is time and energy intensive, but can be optimized leveraging the reduction in features number yielded by the feature selection. In fact, a feature from a specific frequency band, necessitates only its corresponding row in the spectrogram. Therefore, the matrix multiplication can be performed only for the rows that are needed for a feature, reducing one dimension of the Mel Frequency Matrix. This saves both energy and time for the computations, and memory for storing the matrix.

As an embedded harware use-case to deploy our application, we select an Nordic nRF5340-DK, featuring and ARM Cortex-M33 processor for computation, 1 MB of flash memory, 448 KB of utilizable RAM memory, and a maximum frequency of 128 MHz.
The nRF5340-DK features two pins for sensing the current drawn only by the system-on-chip, allowing to obtain a precise and isolated measure.
All experiments are performed utilizing this microcontroller and measuring execution time and current with an Otii Arc Pro device from Quoitech, synchronizing via GPIO.
In particular, the GPIO line can be toggled via software to synchronize with the measurement, so to isolate the desired algorithm to measure.

\subsection{Summary of Experiments}

The experiments performed in this study are listed in Table \ref{tab:experiment_list}. The baseline audio and kinematic models are trained according to the procedure described in Section \ref{exp_setup/ml_test_train}. Then, subsequent optimizations for each model are performed in the order in which they appear in the table. The final, optimized audio and kinematic models are then used in the multimodal model for further optimizations.

\begin{table}[]
\caption{List of performed experiments}
\centering
\begin{tabular}{@{}|l|c|l|@{}}
\hline
Experiment                                                                                                                                                                  & \multicolumn{1}{l|}{Description}                                        & \multicolumn{1}{l|}{Results} \\ \hline \hline
\multicolumn{1}{|l|}{\textbf{Signal selection}} & Section \ref{methodology:sub:hyperparameters:signals} & Section \ref{results:sensor_selection}                     \\ \hline
\textbf{Audio model}                                                                                                                                                        &                                                                         &                              \\
\begin{tabular}[c]{@{}l@{}}Baseline model\\ Sampling frequency selection\\ Window length optimization\\ MFCC vs. Mel spec. features\\ Feature number selection\end{tabular} & \begin{tabular}[c]{@{}l@{}}Section \ref{exp_setup/ml_test_train}\\ Section \ref{methodology:sub:hyperparameters:sampling}\\ Section \ref{methodology:sub:hyperparameter:window_length}\\ Section \ref{methodology:sub:hyperparameters:feature_ext}\\ Section \ref{methodology:sub:hyperparameters:feature_sel}\end{tabular} & Section \ref{results:audio_opt} \\ \hline
\textbf{Kinematic model}                                                                                                                                                    &                                                                         &                              \\
\begin{tabular}[c]{@{}l@{}}Baseline model\\ Window length optimization\end{tabular}                                                                                         & \begin{tabular}[c]{@{}l@{}}Section \ref{exp_setup/ml_test_train}\\ Section \ref{methodology:sub:hyperparameter:window_length}\end{tabular} & Section \ref{results:kinematic_opt} \\ \hline
\textbf{Cough-E}                                                                                                                                                            &                                                                         &                              \\
\begin{tabular}[c]{@{}l@{}}Rerun vs. no-rerun models\\ N\_windows\_max optimization\\ Model decision thresholds\end{tabular}                                                & Section \ref{methodology:sub:multimodal} & Section \ref{results:multimodal_opt} \\ \hline
\end{tabular}
\label{tab:experiment_list}
\end{table}


\section{Results}
\label{results}

\subsection{Edge-AI hyperparameter co-optimization}

\subsubsection{Sensor selection}
\label{results:sensor_selection}

\begin{table}[]
\caption{ML performance for single-sensor models}
\centering
\begin{tabular}{|l|l|}
\hline
Sensor             & CV F-1 Average $\pm$ St. Dev \\ \hline
Kinematic                & 0.646 $\pm$ 0.110              \\ \hline
Body-facing mic    & 0.702 $\pm$ 0.132              \\ \hline
Outward-facing mic & \textbf{0.774 $\pm$ 0.999}     \\ \hline
\end{tabular}
\label{tab:sensor_selection}
\end{table}

Table \ref{tab:sensor_selection} reports the average event-based F-1 score across 15 LOSO-CV folds of models trained using only one signal. The outward-facing microphone has the highest cough event detection performance; the average F-1 scores of the body-facing microphone and kinematic models are respectively 9.3\% and 16.5\% lower relative to the outward-facing microphone. The kinematic model exhibits a 7.9\% reduction in performance realtive to the body-facing microphone.

\subsubsection{Audio model optimization}
\label{results:audio_opt}

The final CV F-1 score of the audio model downsampled at different frequencies is displayed in Table \ref{result:tab:frequency_comparison}. The highest performance occurs at 8 kHz, at which the model exhibits a relative F-1 score 1.67\% higher than that of the original 16 kHz. The model trained on 4 kHz signals results in a CV F-1 score drop of 1.39\% relative to 8 kHz. The audio signals are downsampled to 8 kHz for all further analysis, thus halving the RAM overhead.

\begin{table}[]
\caption{CV F-1 score for different sampling frequencies}
\label{result:tab:frequency_comparison}
\centering
\begin{tabular}{|l|l|}
\hline
Sampling frequency (kHz) & CV F-1 Average $\pm$ St. dev. \\ \hline
16                       & 0.774 $\pm$ 0.999               \\ \hline
\textbf{8}                        & \textbf{0.787} $\pm$ \textbf{0.114}               \\ \hline
4                        & 0.776 $\pm$ 0.105               \\ \hline
\end{tabular}
\end{table}

\begin{figure}[!t]
\centerline{\includegraphics[width=0.9\columnwidth]{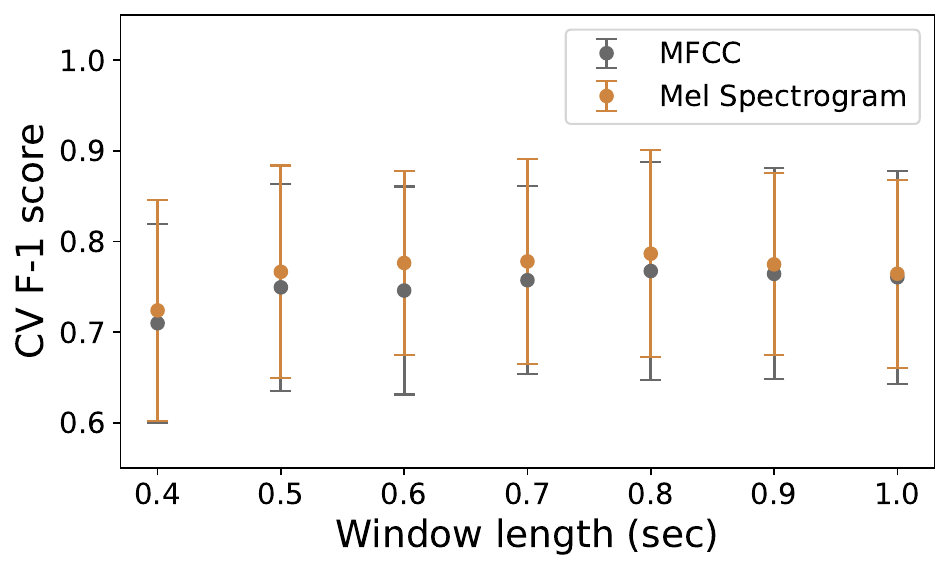}}
\caption{Comparison of CV F-1 score of the audio model using features from the Mel spectrogram and MFCCs at varying window lengths.}
\label{result:fig:mfcc_vs_mel}
\end{figure}

Next, the downsampled signal is used to extract two sets of features, one using the Mel spectral bands and the other using the MFCCs, along with the other audio features listed in Table \ref{tab:extracted_features}. This is performed at varying window lengths, and the resulting CV F-1 scores are displayed in Fig. \ref{result:fig:mfcc_vs_mel}. We first observe that the model with the highest CV F-1 score uses Mel spectrogram features and a window length of 0.8 s.

\begin{figure}[!t]
\centerline{\includegraphics[width=\columnwidth]{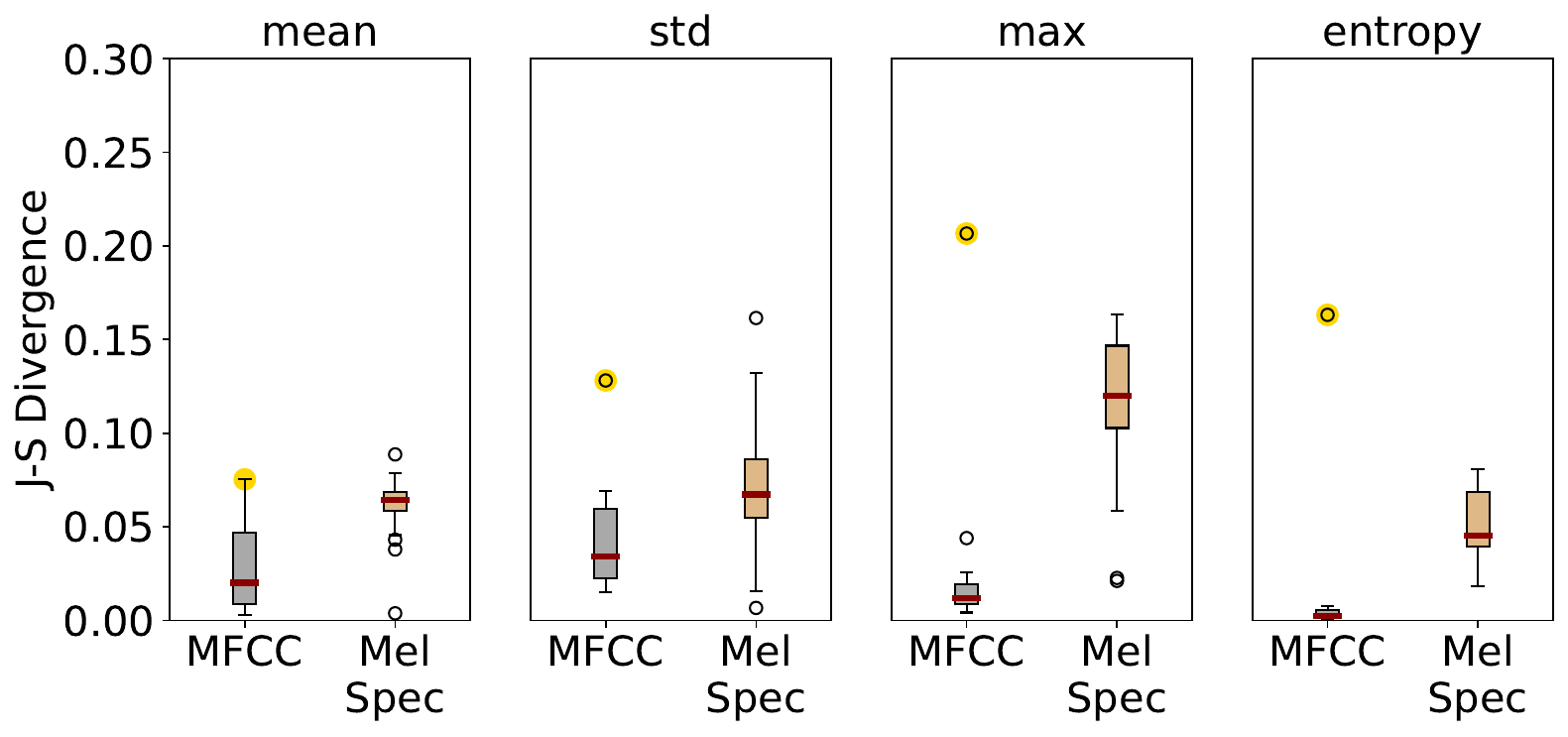}}
\caption{Feature separability between MFCC and Mel Spectrogram using the Jensen-Shannon Divergence. The feature separability of the zero component of the MFCC is highlighted.}
\label{result:fig:feature_separability}
\end{figure}

Fig. \ref{result:fig:mfcc_vs_mel} shows that across all window lengths, the features extracted from the Mel spectrogram have a higher average performance across the CV folds and a comparable standard deviation.
To explain this phenomenon, we quantify the separability between cough and non-cough features across the entire training dataset for the MFCC and Mel spectrogram. Fig. \ref{result:fig:feature_separability} depicts the Jensen-Shannon Divergence, which is a measure of separability between two classes, for each type of extracted feature.
This figure shows that on average, the Mel spectrogram features better distinguish between cough and non-cough windows than MFCC-based features. However, the MFCC extracts one highly separable feature, its zero component, which has a higher separability than any of the Mel spectrogram features for the standard deviation, maximum, and entropy features. 

To further analyze the trade-offs between using Mel spectrogram or MFCC-based features, we analyze the embedded performance metrics of each type of implementation, shown in Table \ref{results:tab:mfcc_mel_times}. 
The computation of MFCC coefficients is computationally demanding, as detailed in Section \ref{methodology:sub:hyperparameters:feature_ext}. By applying raw calculations, the execution time can be prohibitive, requiring 0.929 sec for a 0.5 sec real-time window. The main bottleneck in our case is the DCT step, yielding 16'384 cosine calculations.
To obtain a deployable version, we encoded pre-computed cosine values into a LUT occupying 64kB of memory. 
Instead, the adoption of the Mel spectrogram benefits embedded metrics, requiring a comparable compute time with respect to the MFCC with cosine LUT, but without any need for memory-stored values. Overall, this optimization yields an 20x energy decrease.


\begin{table}[]
\centering
\begin{tabular}{|clc|c|c|c|}
\hline
\multicolumn{3}{|c|}{\textbf{Feature}}                         & \textbf{\begin{tabular}[c]{@{}c@{}}Time\\ {[}ms{]}\end{tabular}} & \textbf{\begin{tabular}[c]{@{}c@{}}Memory\\ {[}KB{]}\end{tabular}} & \textbf{\begin{tabular}[c]{@{}c@{}}Energy\\ {[}$\mu$Wh{]}\end{tabular}} \\ \hline \hline
\multicolumn{2}{|c|}{\multirow{2}{*}{MFCC}} & no LUT           & 929                                                              & -                                                                  & 5.920                                                                \\ \cline{3-6} 
\multicolumn{2}{|c|}{}                      & with LUT         & 54                                                               & 64                                                                 & 0.337                                                                \\ \hline
\multicolumn{3}{|c|}{Mel spectrogram}                          & 49                                                               & -                                                                  & 0.296                                                                \\ \hline
\end{tabular}
\caption{Comparison of MFCC (with and without Look-Up Table for DCT) and Mel spectrogram.}
\label{results:tab:mfcc_mel_times}
\end{table}

\begin{figure}[!t]
\centerline{\includegraphics[width=0.8\columnwidth]{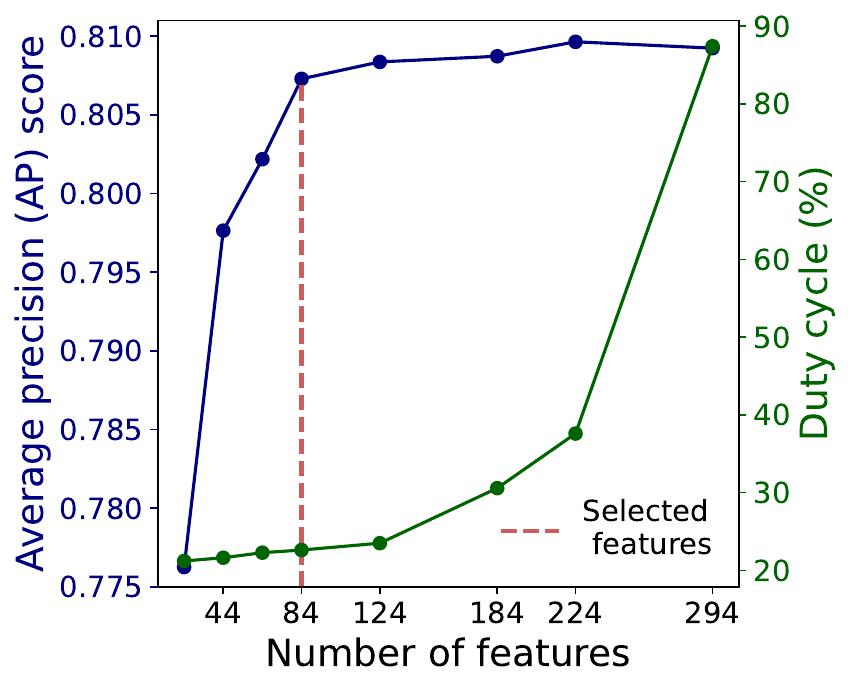}}
\caption{Average Precision and duty-cycle compared to the number of features used by the audio classifier.}
\label{results:fig:feature_elimination}
\end{figure}

Next, we perform RFECV to analyze the effects of feature elimination on runtime and ML performance. Fig. \ref{results:fig:feature_elimination} shows both the duty cycle of the classifier on the embedded platform, as well as the Average Precision (AP) of the model for each number of features. RFECV selects features to remove based on changes to AP. The knee-point of the AP curve occurs when 84 features are selected. It is an operating point with a low duty cycle (23.62\%).

Table \ref{tab:final_classifier_details} displays the implementation details of the final classifier, which is trained using the entire training dataset and used for the final testing phase. This table also displays the five most significant features impacting the classification outcome. 
The Mel spectrogram features, both at low frequencies (component 0) and high frequencies (components 56, 60, and 61) contribute over half of the total feature importance.

\newcolumntype{?}{!{\vrule width 1pt}}

\begin{table}[]
\centering
\caption{Final trained classifier details}
\begin{tabular}{|c?c|c|}
\hline
                                                                                               & \textbf{Audio model}                                                                                                                                           & \textbf{Kinematic model}                                                                                                                           \\ \hline \hline
\begin{tabular}[c]{@{}c@{}}Signals\end{tabular}                                              & \begin{tabular}[c]{@{}c@{}}Outward-facing\\ mic, Gender\end{tabular}                                                                                        & \begin{tabular}[c]{@{}c@{}}Accel X,Z, magnitude\\ Gyro Y,P,R, magnitude\\ Gender, BMI\end{tabular}                                           \\ \hline
Sampling freq.                                                                                          & 8 kHz                                                                                                                                                       & 100 Hz                                                                                                                                     \\ \hline
Window len.                                                                                             & 0.8 s                                                                                                                                                       & 0.5 s                                                                                                                                      \\ \hline
Classifier                                                                                              & XGB                                                                                                                                                         & XGB                                                                                                                                        \\ \hline
No. features                                                                                            & 84                                                                                                                                                          & 36                                                                                                                                         \\ \hline
\begin{tabular}[c]{@{}c@{}}Top 5 most\\ significant \\ features\\ (\% importance)\end{tabular} & \begin{tabular}[c]{@{}c@{}}Mel \#60 max (43.0)\\ Mel \#0 st. dev. (8.7)\\ Mel \#61 st. dev (6.9)\\ Crest factor (3.1)\\ Mel \#56 st. dev (2.9)\end{tabular} & \begin{tabular}[c]{@{}c@{}}Roll LL (11.7)\\ Roll AZC,$\epsilon=1$ (8.5)\\ Accel Z AZC,$\epsilon=0.9$ (8.5)\\ Gender (5.9)\\ BMI (3.7)\end{tabular} \\ \hline
\end{tabular}
\label{tab:final_classifier_details}
\end{table}

\subsubsection{Kinematic model optimization}
\label{results:kinematic_opt}
Due to the lower sampling frequency of the kinematic model, its energy consumption is not as crucial to optimize and therefore design decisions were taken to maximize the LOSO CV F-1 score. The hyperparameters of the model are displayed in Table \ref{tab:final_classifier_details}. Out of the original extracted 106 kinematic features (13 for each signal, and 2 subject information features), the classifier using only 36 features exhibited the highest performance in RFECV. All features extracted using only the accelerometer y-direction signal, which captures side-to-side movement, were eliminated. 

\subsection{Cough-E model optimization}
\label{results:multimodal_opt}
To optimize the multimodal Cough-E classifier, we first analyze the effects of running the model in the \textit{re-run} versus \textit{no-rerun} modes, shown in Table \ref{tab:simulator_test_results}.
The impact on runtime and energy is estimated in terms of percentage of audio and kinematic model executions.
\textit{no-rerun} mode exhibits the fewest audio executions (20.25\%) at the cost of the lowest F-1 score (0.692). \textit{rerun} mode instead provides a slightly more computationally-complex execution, with 4.96\% more audio model runs for an F-1 score 17.91\% higher relative to \textit{no-rerun}. Moreover, relatively to continuously running only the audio model, its CV F-1 score increases by 5.83\%.


\begin{table}[]
\centering
\begin{tabular}{|c|c|c|c|}
                \hline
                 & \textbf{Audio} & \multicolumn{2}{|c|}{\textbf{Cough-E}} \\
                \cline{3-4}
                & \textbf{only} & \textbf{no-rerun} & \textbf{rerun} \\
                \hline\hline
                \textbf{\% Audio} & 100 & 20.25 & 25.21 \\
                \hline
                \textbf{\% Kinematic} & 0 & 79.75 & 74.79 \\
                \hline
                \textbf{Avg. CV F-1} & 0.771 & 0.692 & 0.816 \\
                \hline
\end{tabular}
\caption{Comparison of performance of the base and multimodal configuration, \textit{no-rerun} and \textit{rerun} case.}
\label{tab:simulator_test_results}
\end{table}

\begin{figure}[!t]
\centerline{\includegraphics[width=0.8\columnwidth]{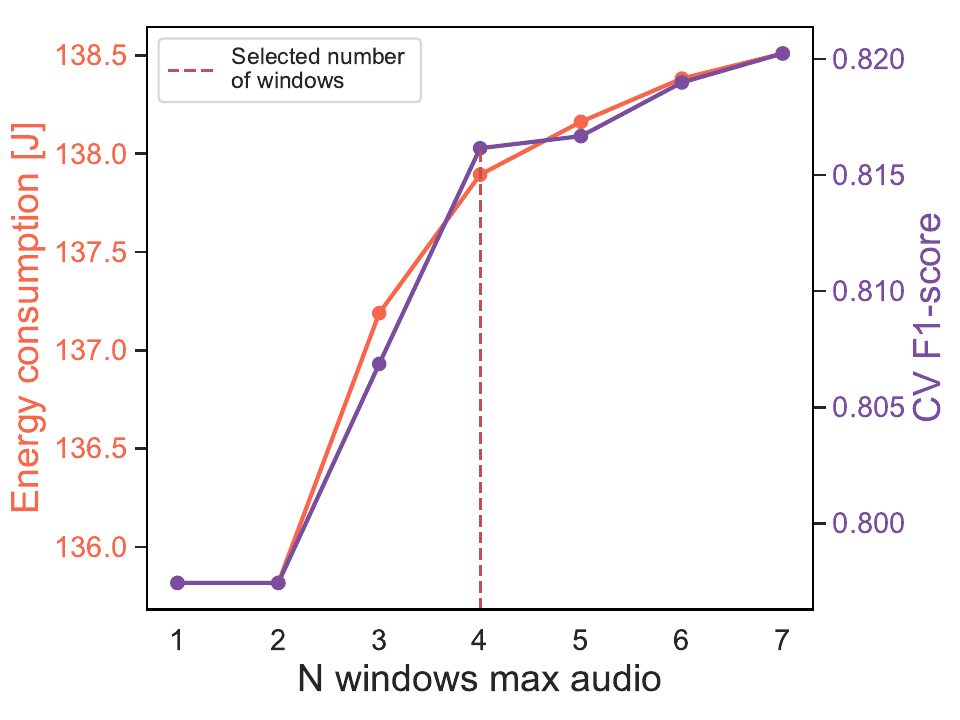}}
\caption{Effects on CV F-1 score and energy of $N^{windows}_{max}$.}
\label{result:fig:n_windows_audio_optim}
\end{figure}

Next, we perform an exploration of $N^{windows}_{max}$, comparing CV F1-score and energy consumption, simulated on the train set, for values spanning from 1 to 7, as reported in Figure \ref{result:fig:n_windows_audio_optim}.
From this exploration, we choose a final $N^{windows}_{max}$ value equal to 4, representing a suitable knee point for the comparison curves.

\begin{figure}[!t]
\centerline{\includegraphics[width=0.9\columnwidth]{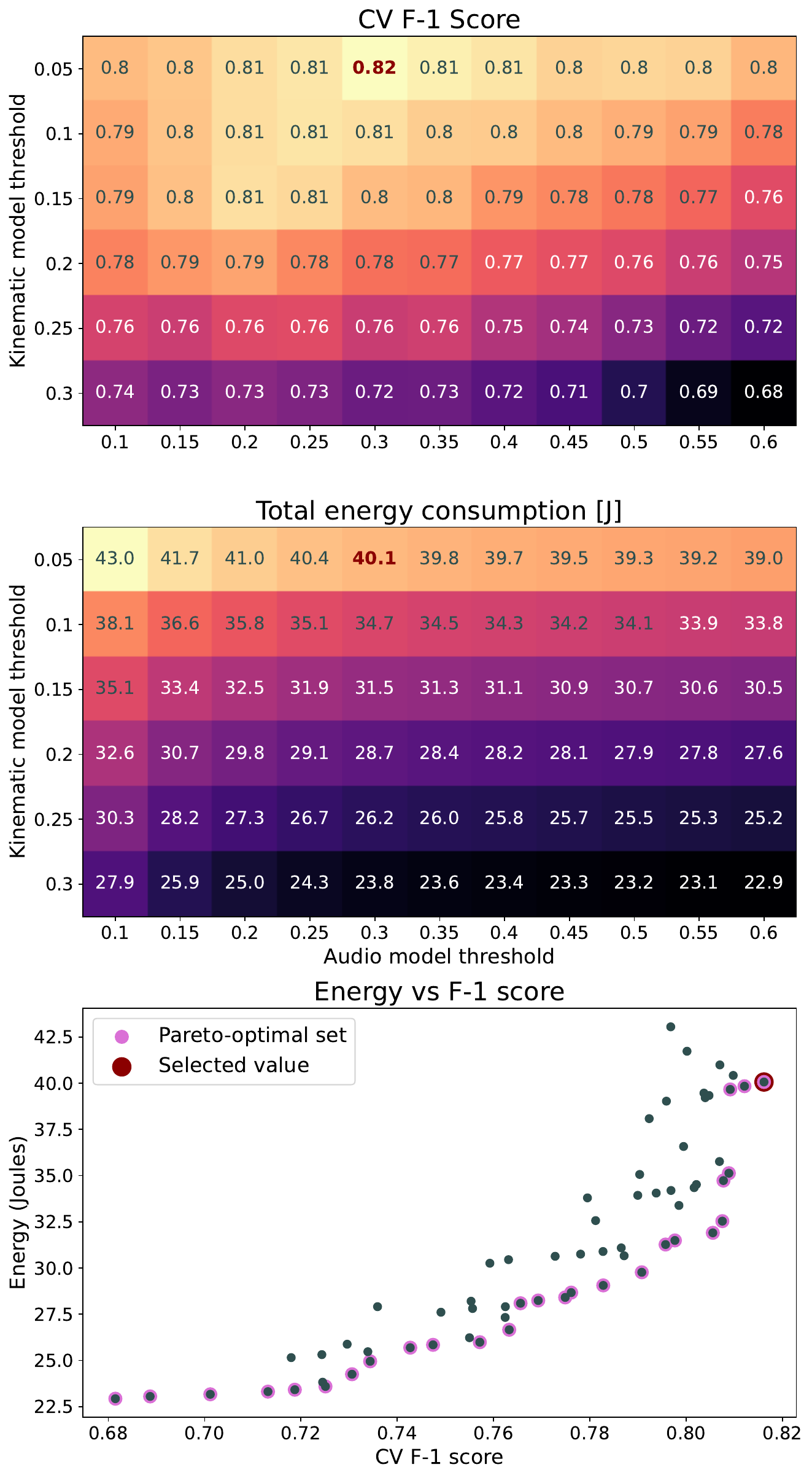}}
\caption{Cross-validation F-1 score and total energy consumption of the multi-modal modal at different decision thresholds.}
\label{result:fig:multimodal_threshold_opt}
\end{figure}

The final hyperparameters that are co-optimized are the decision thresholds of the audio and kinematic classifiers that influence which segments are counted as cough-positive and consequently control switching between models. As these parameters are highly interdependent, they are co-optimized to form the top graph in Fig. \ref{result:fig:multimodal_threshold_opt}. The optimal average CV F-1 score of 0.82 occurs at the lowest kinematic threshold of 0.05, and a higher audio threshold of 0.3. In parallel, the energy consumption resulting from the combination of thresholds is reported in the central part of Fig. \ref{result:fig:multimodal_threshold_opt}. We observe an inverse correlation between energy consumption and both decision thresholds. To select a final combination of decision thresholds for testing the multimodal model, all combinations of F-1 score and energy are plotted in the lower part of Fig \ref{result:fig:multimodal_threshold_opt}. The Pareto frontier of points illustrates the ones that optimize the energy-accuracy trade-off. It happens that the combination yielding the highest CV F-1 score is on this front, so this value is selected to maintain a high ML performance.

\subsection{Noise robustness analysis}
Table \ref{tab:final_cv_results_ml} displays the final CV results, including the average and standard deviation of each metric across the LOSO CV folds, of each model with its finalized hyperparameters. We see that Cough-E performs best in CV, with the highest average F-1 score, lowest FP/hr, and lowest standard deviation across both metrics. The low F-1 score of the kinematic model is mainly due to a high number of FPs, manifested through the high FP/hr and low PR. 

\begin{table}[]
\centering
\caption{Final CV ML performance results}
\begin{tabular}{|l|l|l|l|l|}
\hline
Model      & SE                                                      & PR                                                     & F-1                                                    & FP/Hr                                                \\ \hline
Kinematic        & \begin{tabular}[c]{@{}l@{}}0.82 $\pm$ \\ 0.14\end{tabular} & \begin{tabular}[c]{@{}l@{}}0.57 $\pm$\\ 0.16\end{tabular} & \begin{tabular}[c]{@{}l@{}}0.65 $\pm$\\ 0.11\end{tabular} & \begin{tabular}[c]{@{}l@{}}368 $\pm$\\ 289\end{tabular} \\ \hline
Audio      & \begin{tabular}[c]{@{}l@{}}0.79 $\pm$\\ 0.21\end{tabular}  & \begin{tabular}[c]{@{}l@{}}0.80 $\pm$\\ 0.08\end{tabular} & \begin{tabular}[c]{@{}l@{}}0.77 $\pm$\\ 0.12\end{tabular} & \begin{tabular}[c]{@{}l@{}}188 $\pm$\\ 101\end{tabular} \\ \hline
Cough-E & \begin{tabular}[c]{@{}l@{}}0.80 $\pm$\\ 0.19\end{tabular}   & \begin{tabular}[c]{@{}l@{}}0.87 $\pm$\\ 0.08\end{tabular} & \begin{tabular}[c]{@{}l@{}}0.82 $\pm$\\ 0.11\end{tabular} & \begin{tabular}[c]{@{}l@{}}78 $\pm$\\ 59\end{tabular}   \\ \hline
\end{tabular}
\label{tab:final_cv_results_ml}
\end{table}

\begin{figure}[!t]
\centerline{\includegraphics[width=\columnwidth]{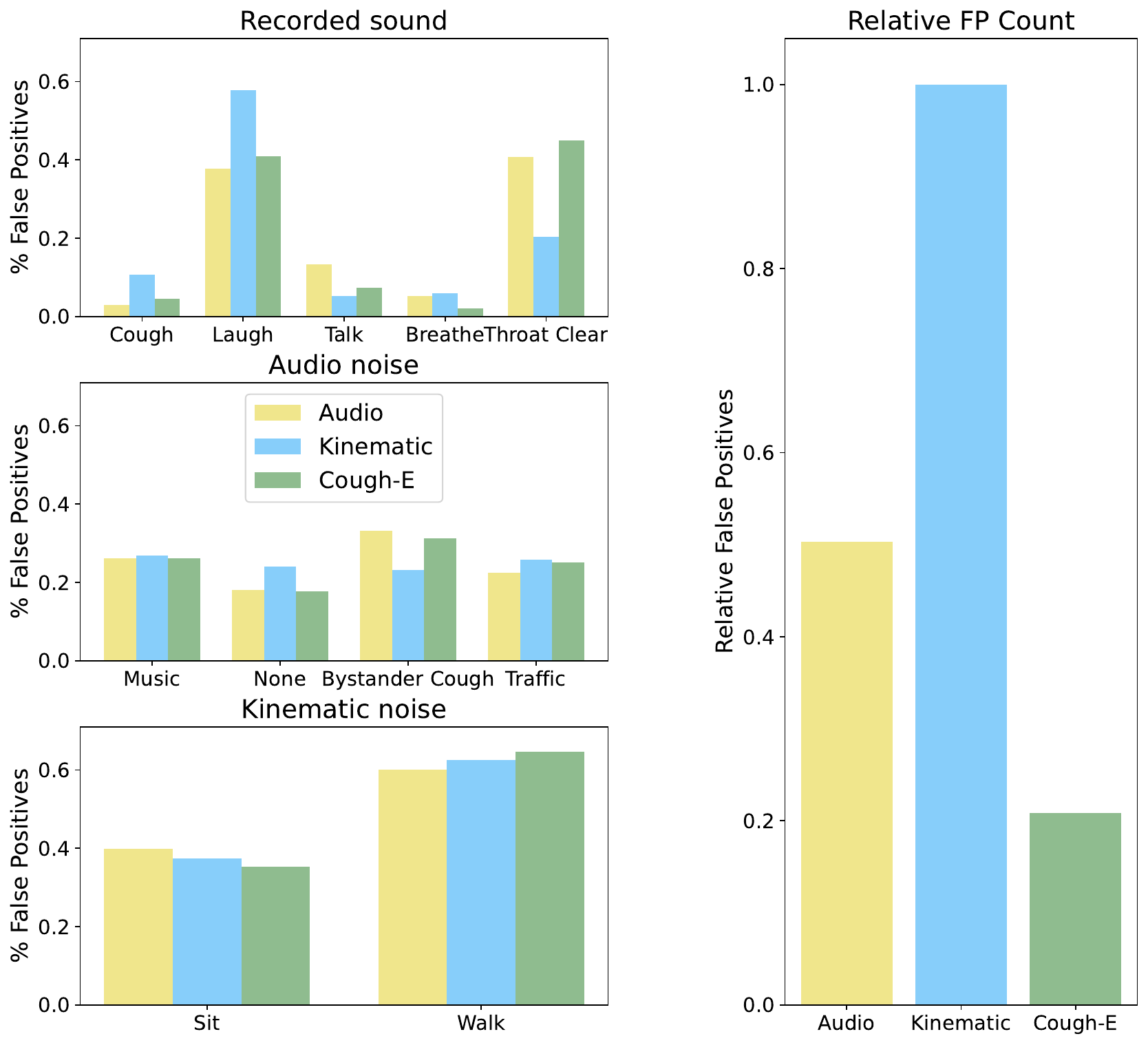}}
\caption{Cross-validation FPs of each model configuration for each class of sound, audio, and kinematic noise recording.}
\label{result:fig:fp_distribution}
\end{figure}

Overall, we observe high FP/hr metrics across all models, which is due to the composition of the dataset. As described in Section \ref{exp_setup/dataset}, the data contain four times more data in audio noise scenarios than silence, as well as 50\% kinematic noise. Furthermore, in 20\% of the recordings, less sounds like laughter and throat clearing are made, which is not reflective of a real-life scenario. To further analyze how each model behaves in the presence of each parasitic sound and noise scenario, the percentage of false positives (FPs) occurring in each scenario for the three models is shown in Fig. \ref{result:fig:fp_distribution}.


The first part of Fig. \ref{result:fig:fp_distribution} shows that in the most common everyday scenarios, where the person is talking, breathing, or coughing, few FPs occur. In contrast, most FPs occur during laughing and throat clearing. Laughing in particular accounts for 57.7\%, 37.8\%, and 40.9\% of the FPs in the kinematic, audio, and Cough-E models, respectively.

The middle section of Fig. \ref{result:fig:fp_distribution} shows the FP distribution in different audio noise scenarios. While the FPs of the kinematic model are relatively constant across audio noise scenarios (st. dev$=$1.4\%), the audio model performance fluctuates (st. dev$=$5.5\%), exhibiting the most FPs when a bystander is coughing next to the subject. Cough-E performance changes slightly less than the audio model across noises (st. dev$=$4.8\%). The lower part of the figure shows the effects of kinematic noise on each classifier. When going from sitting to walking, the FP percentage increases by 20.3\%, 25.1\%, and 29.4\% for the audio, kinematic, and Cough-E models, respectively. Finally, the right of Fig. \ref{result:fig:fp_distribution} shows the relative number of FPs normalized to the sum of FPs of the IMU model. We observe that Cough-E exhibits 2.4x and 4.8x fewer FPs than the audio and kinematic models, respectively.

\subsection{Final testing results}

Table~\ref{tab:mem_requirements} shows the memory requirements (divided into peak RAM usage and LUT occupation) of the final embedded application.
The intensive audio classifier dictates the 191.0 kB overall peak RAM usage. Analogously, the audio classifier occupies the main memory for LUTs, 136.73 kB more than the kinematic one and 66.31\% of the overall occupation. 

The final ML performance and energy consumption results on the private test subjects are shown in Table \ref{tab:FINAL_TEST_WOOHOO}. The energy consumption is measured by running on the whole test set. The un-optimized audio model is not suitable for the selected microcontrollers, exhibiting a peak RAM usage of 473.2 kB, around 10 kB higher than the available memory. Therefore, its energy consumption is not measured. 

First, we observe that our audio model optimizations result in a relative 3.65\% drop in F-1 score.
Next, among the optimized models, the audio exhibits the highest SE, F-1, and energy, with Cough-E and kinematic models exhibiting 1.26\% and 32.91\% lower relative F-1s, and 70.56\% and 92.53\% energy saving, respectively. 


\begin{table}[]
\centering
\begin{tabular}{|l|l|l|}
\hline
\textbf{Algorithm} & \textbf{Peak RAM {[}kB{]}} & \textbf{LUTs {[}kB{]}} \\ \hline \hline
Audio              & 191.00                     & 277.85                 \\ \hline
Kinematic          & 3.10                       & 141.12                 \\ \hline
Cough-E          & 191.0                      & 418.97                 \\ \hline
\end{tabular}
\caption{Memory requirements of the application, detailed for feature extraction and inference steps per model.}
\label{tab:mem_requirements}
\end{table}

\begin{table}[]
\centering
\caption{Final testing results}
\begin{tabular}{|c|c|c|c|c|cc|}
\hline
\multirow{2}{*}{\textbf{Model}} & \multirow{2}{*}{\textbf{SE}} & \multirow{2}{*}{\textbf{PR}} & \multirow{2}{*}{\textbf{F1-score}} & \multirow{2}{*}{\textbf{FP/Hr}} & \multicolumn{2}{c|}{\textbf{Energy}}                       \\ \cline{6-7} 
                                &                              &                              &                                    &                                 & \multicolumn{1}{c|}{\textbf{{[}J{]}}} & \textbf{saving} \\ \hline \hline
Unopt. Audio                    & 0.87                         & 0.78                         & 0.82                               & 160                             & \multicolumn{1}{c|}{-}                & -                  \\ \hline
Opt. Audio                      & 0.8                          & 0.78                         & 0.79                               & 149                             & \multicolumn{1}{c|}{36.99}            & 0.0\%               \\ \hline
Kinematic                       & 0.7                          & 0.43                         & 0.53                               & 304                             & \multicolumn{1}{c|}{2.76}             & 92.53\%              \\ \hline
Cough-E                      & 0.71                         & 0.86                         & 0.78                               & 49                              & \multicolumn{1}{c|}{10.89}            & 70.56\%              \\ \hline
\end{tabular}
\label{tab:FINAL_TEST_WOOHOO}
\end{table}


\section{Discussion}

This work tackles the energy-efficient design of a real-time cough monitor for edge devices by proposing a HW-aware methodology, successfully achieving a multimodal classifier suitable for low-power wearable devices. 

\textit{Audio classifier}: during development, we started by analyzing and comparing outward- and inward-facing audio signals, and evaluating classifiers based on them.
The outward-facing audio model outperformed the body-facing one, probably due to motion artefacts. 
The proximity of the body-facing microphone to the skin may introduce both kinematic and audio noise in key frequency ranges.
Therefore, even though the outward-facing microphone is more exposed to outside noise, it is more performant. 
Using a single audio sensor also reduces memory and energy needs by avoiding the storing and processing of additional data.

The results of the audio frequency optimization are consistent with visual analyses of cough spectrograms, in which most of the spectral energy lies below 4 kHz.
Furthermore, the higher performance of the 8 kHz model compared to the 16 kHz one can be due to the finer granularity of Mel frequency bands at the reduced sampling frequency, leading to more robust features. 
From an embedded perspective, halving the sampling rate means halving the data to store and process, thus reducing memory and computational needs which reduces energy consumption.

The initial set of audio features is taken from prior work \cite{orlandic_coughvid_2021, orlandic2023multimodal}. Moreover, this work further decomposes and analyzes the most computationally complex features: the Mel Frequency Cepstral Coefficients (MFCCs). The goal of this set of features is to extract spectral information on a logarithmic frequency scale similar to the way the human ear distinguishes sounds \cite{davis_comparison_1980}. It has been applied extensively, particularly to classical ML in acoustic applications, showing an important performance impact \cite{abdul2022mel}.
As observed in the maximum and entropy features in Fig. \ref{result:fig:feature_separability}, the MFCC compresses the spectral information into one highly separable feature using the DCT computation. 
However, this results in high processing and memory overhead (cf. Table \ref{results:tab:mfcc_mel_times}).
By using the Mel spectrogram features, we improve the model's performance and reduce energy and processing overhead.

As the model based on Mel spectrogram features achieves the highest ML performance and lowest runtime, these features are used in all further experiments. However, as shown in Table \ref{tab:extracted_features} the Mel spectrogram is a decompressed version of the MFCCs and therefore contains more features. This makes the feature elimination step more important to reduce processing and remove noisy features that diminish model performance.  
As shown in Fig. \ref{results:fig:feature_elimination}, using more features yields increased model's performance and processing overhead.
In particular, the duty cycle of the application is marginally affected for low numbers of features but increases rapidly starting from 183 features. This is due to the Energy Envelope Peak Detection (EEPD) features, required only in models with a higher number of features. EEPDs require time consuming operations such as multiple filtering, vector normalization, and peak identification, repeated for each feature of this type. Such features that cause significant processing overhead and marginal performance gains are thus eliminated by RFECV.

In addition to all of the aforementioned optimizations, our choice of using an XGB classifier rather than large neural networks improves ease of use on embedded devices.
Moreover, its explainability allows us to observe that the most important features come from the Mel spectral domain, suggesting that the model is learning from features that mimic the function of the human ear to distinguish cough from non-cough spectra.

\textit{Kinematic classifier}: the kinematic model goes through fewer optimization steps due to the lower bandwidth and consequent lower processing overhead of the signals.
The optimal window length, 0.5 sec, was shorter than that of the audio model, perhaps because the chest accelerations due to coughing do not last as long as the audio events, as can be observed in Fig. \ref{fig:cough_delineation}. Table \ref{tab:final_classifier_details} shows that the final kinematic model did not use any features from the side-to-side direction. 
A cough typically results in sagittal acceleration with some up-and-down movement.  
The two most important features came from the roll signal, which captures rotation around the body's vertical axis. Perhaps this is due to subjects' reflex to cough into an elbow to the side of the body.

\textit{Cough-E multimodal classifier}: after having optimized audio and kinematic classifiers, the analyses focused on the hyperparameters of the multimodal model execution.
This illustrates key trade-offs in when to use the complex audio classifier versus the more lightweight kinematic one. Table \ref{tab:simulator_test_results} shows that using the \textit{rerun} mode reduces the number of audio model executions by nearly 75\% with a relative 5.83\% boost in ML performance, in contrast with the \textit{no-rerun} mode that causes around 80\% and 10.37\% relative drop respectively. 
\textit{no-rerun} mode is limited by the kinematic model being the only one processing the first segment of a positive classification. The first cough event may only be included in the kinematic window that triggers the audio classifier. It could be counted from the kinematic model's output, but this would yield a high number of FPs, due to the high sensitivity of the triggering model, thus decreasing the performances.
This shows the utility of the audio model in reassessing kinematic-based detection.
Next, when optimizing $N^{windows}_{max}$ in Fig \ref{result:fig:n_windows_audio_optim}, we showed that low values prevent long audio model runtimes. This marginally affects the performance but saves computational energy, allowing the more efficient kinematic model to run more often. Similarly, when optimizing the decision thresholds of the kinematic and audio models in Fig. \ref{result:fig:multimodal_threshold_opt}, we observe that low kinematic thresholds and moderate audio ones maximize CV F-1 score. This is because low kinematic thresholds maintain a high SE but many FPs, while the audio model then validates the output with a higher threshold. From an energy standpoint, a lower kinematic threshold leads to higher energy consumption because the energy-consuming audio model is activated more often. A higher audio threshold, on the other hand, saves energy because it means that the model switches back to the more efficient kinematic model sooner. Using Pareto analysis, we select an optimal value to optimize the model-switching activity in terms of energy and ML performance. 

In addition to its lower energy consumption, Cough-E also exhibits noise discrimination capabilities, shown in Fig. \ref{result:fig:fp_distribution}. The FP/hr metric of the multimodal model is lower than both the kinematic and audio models across all noise scenarios. Since the FPs of the audio and kinematic classifiers tend to occur during different types of noise, the multimodal model exhibits fewer FPs overall and consequently yields a higher PR than both models. While the audio model is less resistant to motion artifacts, it does exhibit more FPs, perhaps because of the sounds of the subject walking or consequent movement of the microphone along the subject's clothing. As a result, the multimodal model only fails when both the audio and kinematic models fail.

In the CV results shown in Table \ref{tab:final_cv_results_ml}, Cough-E had the highest SE and F-1. However, between the mean CV metrics and the final test in Table \ref{tab:FINAL_TEST_WOOHOO}, we observe 14.63\% and 11.25\% drops in SE of the kinematic and multimodal Cough-E classifiers, respectively, while the SE of the audio model stays roughly the same. This could be explained by over-fitting of the kinematic model, whereas the audio model was pre-trained with the COUGHVID dataset and was therefore less prone to over-fitting. Due to the loss in kinematic SE, Cough-E does not detect many coughs in the first place and therefore its F-1 score drops. However, Cough-E still exhibits the lowest FP/hr, precisely 3x and 14.9x lower than those of the audio and kinematic models, respectively.
Table \ref{tab:FINAL_TEST_WOOHOO} shows how our methodology succeeds in achieving an efficient edge-AI execution. Cough-E exhibits a balanced trade-off between performance and energy consumption. This is due to the optimized collaboration of the kinematic model, yielding the lowest consumption but also the lowest F1-score, and the audio model. The latter features a 3.65\% F1-score drop, relative to the unoptimized one, due to the HW-aware optimizations but is executable on the edge, although with a high energy consumption.
Measurements are not performed for the unoptimized audio model due to its high memory consumption that prevents its embedded implementation, further proving the need for our methodology to enable efficient edge AI cough detection algorithm design.

The literature works presented in Section \ref{sec:introduction} exhibit high ML performances, often reported in terms of SE and PR. However, they are typically measured on a segment level, prone to fail in capturing the precise number of cough events. Our work overcomes this limitation by detecting coughs on an event level.
Additionally, few works reported execution on the edge, mainly targeting onerous algorithms for offline computation on collected data.

Although the audio model is pre-trained with the COUGHVID dataset, our results are limited to a forced-coughs dataset. Therefore, future research would be to apply our methodology to clinical data collected in hospitals.
Furthermore, an interesting analysis could be done on the use of raw gyroscope signals, instead of the derived angles, exploring their energy and performance impact.

\section{Conclusion}

In this work, we have introduced a novel methodology for automated, energy-efficient, and edge AI cough monitoring algorithms, analyzing with fine granularity all stages of development of the pipeline.
Notably, we developed a complex audio classifier and a light kinematic-based one, while assessing the impact of design hyperparameters on performance and energy. Performances were evaluated using a novel clinically significant event-based metric, and energy was measured on a real embedded platform commonly found in wearable devices. 
We demonstrated the efficiency of utilizing Mel spectrogram features instead of MFCC ones, achieving a 20x energy reduction together with an increase in performance.
Furthermore, our optimizations of the sampling rate, window length, and number of implemented features enabled the execution of the audio-based model on a resource-constrained embedded platform.
We then studied the multimodal integration of the kinematic classifier to trigger the execution of the audio-based one. 
This involved analyzing their interaction and the balance between performance and energy by limiting the audio model executions.
This culminated in an edge-AI, multimodal cough detection algorithm, deployed on a low-power microcontroller with a 70.56\% energy reduction at the cost of only 1.26\% F1-score decrease, compared to constantly running the robust audio model.
This work has demonstrated the importance of the edge AI deployment of cough monitors, achievable through our HW-aware methodology, paving the way to the 24-hour privacy-preserving monitoring of patients with chronic cough.






\begin{small}
    \printbibliography
\end{small}




\end{document}